\documentclass[namedreferences]{solarphysics}

\usepackage[hyperref,optionalrh,natbib]{spr-sola-addons} 
\usepackage{graphicx}        
\usepackage{amssymb}        
\usepackage{color}           

\newcommand{\etal}{{\it et al.}}

\newcommand{\arcsec}{''}
\newcommand{\degree}{^{\circ}}

\newcommand{\kms}{km s$^{-1}$} 
\newcommand{\kmss}{km s$^{-2}$} 
\newcommand{\Rsun}{$R_{\odot}$}

\newcommand{\eg}{\textit{e.g.}}
\newcommand{\ie}{\textit{i.e.}}

\newcommand{\BE}{\begin{equation}}
\newcommand{\EE}{\end{equation}}
\newcommand{\BA}{\begin{eqnarray}}
\newcommand{\EA}{\end{eqnarray}}
 \newcommand{\fig}[1]{Figure~\ref{fig:#1}}
 \newcommand{\figs}[2]{Figures~\ref{fig:#1} and \ref{fig:#2}}
 
 \newcommand{\sect}[1]{Section~\ref{sec:#1}}

 \newcommand{\eq}[1]{Equation~(\ref{eq:#1})}

\renewcommand{\vec}[1]{{\mathbfit #1}}

\begin{document}

\begin{article}
\begin{opening}

\title{Moreton and EUV Waves Associated with an X1.0 Flare and CME Ejection\\ {\it Solar Physics}}

\author[addressref={1},email={cfrancile@unsj-cuim.edu.ar}]{\inits{C.N.}\fnm{Carlos}~\lnm{Francile}}
\author[addressref={2},email={flopez@icate-conicet.gob.ar}]{\inits{F.M.}\fnm{Fernando M.}~\lnm{L\'opez}}
\author[addressref={3,4},email={hebe.cremades@frm.utn.edu.ar}]{\inits{M.H.}\fnm{Hebe}~\lnm{Cremades}}
\author[addressref={5,6},corref,email={mandrini@iafe.uba.ar}]{\inits{C.H.}\fnm{Cristina H.}~\lnm{Mandrini}}
\author[addressref={5},email={mluoni@iafe.uba.ar}]{\inits{M.L.}\fnm{Mar\'\i a~Luisa}~\lnm{Luoni}}
\author[addressref={7},email={david.long@ucl.ac.uk}]{\inits{D.M.}\fnm{David M.}~\lnm{Long}}

\runningauthor{C.N.~Francile \etal}
\runningtitle{Comparative Analysis of a Moreton and EUV Wave}

\address[id={1}]{Observatorio Astron\'omico F\'elix Aguilar (OAFA), UNSJ, San Juan, Argentina}
\address[id={2}]{Instituto de Ciencias Astron\'omicas, de la Tierra y del Espacio (ICATE), CONICET, San Juan, Argentina}
\address[id={3}]{Universidad Tecnol\'ogica Nacional, Facultad Regional Mendoza, Mendoza, Argentina}
\address[id={4}]{Consejo Nacional de Investigaciones Cient\'\i ficas y T\'ecnicas (CONICET), Mendoza, Argentina}
\address[id={5}]{Instituto de Astronom\'\i a y F\'\i sica del Espacio (IAFE), CONICET-UBA, Buenos Aires, Argentina}
\address[id={6}]{Facultad de Ciencias Exactas y Naturales (FCEN), UBA, Buenos Aires, Argentina}
\address[id={7}]{UCL-Mullard Space Science Laboratory, Holmbury St Mary, Dorking, Surrey, RH5 6NT, UK}

\begin{abstract}
A Moreton wave was detected in active region (AR) 12017 on 29 March 2014 with very high cadence with the {\it Halpha Solar Telescope for Argentina} (HASTA)  in association with an X1.0 flare (SOL2014-03-29T17:48). Several other phenomena took place in connection with this event, such as low coronal waves and a coronal mass ejection (CME).  
We analyze the association between the Moreton wave and the EUV signatures observed with the {\it Atmospheric Imaging Assembly} onboard the {\it Solar Dynamics Observatory}. These include their low-coronal surface-imprint, and the signatures of the full wave and shock dome propagating outward in the corona. We also study their relation to the white-light CME. We perform a kinematic analysis   
by tracking the wavefronts in several directions. 
This analysis reveals a high-directional dependence of accelerations and speeds determined from data at various wavelengths. We speculate that a region of open magnetic field lines northward of our defined radiant point sets favorable conditions for the propagation of a coronal magnetohydrodynamic shock in this direction. The hypothesis that the Moreton wavefront is produced by a coronal shock-wave that pushes the chromosphere downward
is supported by the high compression ratio in that region. Furthermore, we propose a 3D geometrical model to explain the observed wavefronts as the chromospheric and low-coronal traces of an expanding and outward-traveling bubble intersecting the Sun. The results of the model are in agreement with the coronal shock-wave being generated by a 3D piston that expands at the speed of the associated rising filament. The piston is attributed to the fast ejection of the filament\,--\,CME ensemble, also consistent with the good match between the speed profiles of the low-coronal and white-light shock-waves.
\end{abstract}

\end{opening}

\section{Introduction}
\label{sec:intro}

Coronal mass ejections (CMEs) and flares are the most energetic events that occur in the solar atmosphere. Their effects are detectable not only close to the Sun, but also far into the interplanetary medium; they are key drivers of space-weather conditions.
CMEs and flares can be observed separately or in conjunction, and they are generally associated with large active regions (ARs) with a complex magnetic field structure.
The influence of flares and CMEs on the solar atmosphere includes a wide ariety of phenomena, such as radio bursts, accelerated particle beams, formation of transient coronal holes, shocks, and large-scale propagating perturbations like Moreton and EUV waves among others (\opencite{Benz08}; \opencite{Chen11}; \opencite{Webb12}).

Globally propagating waves as a byproduct of flares were first explained by \citeauthor{Uchida68} (\citeyear{Uchida68}, \citeyear{Uchida73}).  Uchida's magnetohydrodynamic (MHD) model interpreted the disturbances observed in H$\alpha$, which were discovered by \inlinecite{Moreton60}, \inlinecite{Moreton60b}, and \inlinecite{Athay61}. These so-called Moreton waves were observed in association with the most impulsive flares as arc-shaped fronts with a restricted angular span propagating with speeds of 500\,--\,1500 \kms~to distances of up to 400\,Mm. 
In Uchida's model, a coronal fast-mode MHD wave, or eventually a shock, originates within an AR, evolves as a full 3D wave-dome out to the corona, and produces enhanced H$\alpha$ emission when it impacts and sweeps the chromosphere. \citeauthor{Warmuth01} (\citeyear{Warmuth01}, \citeyear{Warmuth04a},
\citeyear{Warmuth04b}) (see also \inlinecite{Warmuth15} and references therein) found that Moreton perturbations slow down as they propagate, while their intensity profiles in H$\alpha$ decrease and broaden as they move away from their source. In their view, and in agreement with Uchida's model, this behavior could be explained by a freely propagating fast-mode MHD shock created by a
large-amplitude single pulse, which might finally decay to an ordinary fast-mode MHD wave. The analysis of numerical simulations by \inlinecite{Vrsnak16} addresses the necessary conditions for an eruption to cause an observable Moreton wave and a coronal shock front, with weaker eruptions producing only coronal and transition region signatures. According to their results, the perturbation evolves as a freely propagating simple wave after the initial eruption-driven phase.

Observational evidence of coronal waves in the EUV range was first reported by \inlinecite{Moses97} and \inlinecite{Thompson98} using data from the {\it Extreme Ultraviolet Imaging Telescope} (EIT: \opencite{Delaboudiniere95}), onboard the {\it Solar and Heliospheric Observatory} (SOHO) satellite.
These so-called EIT waves seen on the solar disk show slower propagating  speeds than Moreton waves, about 200\,--\,400 \kms, and are visible for $\approx$ 45\,--\,60 min after the event that originates them. The EUV fronts propagate almost radially from an AR, with a span of up to $360\degree$, to distances beyond one solar radius. They are usually faint and diffuse and weaken as they propagate. However, they are occasionally seen as an arc-like sharp front, in which case they are called S-waves or brow waves 
(\opencite{Warmuth15} and references therein). Evidence of large-scale wavefronts has been also observed in soft X-rays (see \eg~\opencite{Khan02}; \citeauthor{Narukage02} \citeyear{Narukage02}, \citeyear{Narukage04}; \opencite{Hudson03}). They are observed as arc-shaped emission enhancements, are more homogeneous, and have sharper leading edges than those of EIT waves.

The discrepancy between Moreton- and EIT-wave speeds started a controversy over years about the nature of EIT waves and their role as coronal counterparts of the  observable chromospheric phenomenon (see \inlinecite{Warmuth15} and references therein). This discrepancy led to suggestions that the two phenomena were physically different and propagated independently. \inlinecite{Warmuth01} proposed that the mismatch in speeds might arise because Moreton waves are always seen to decelerate, while the very low temporal cadence of EIT could lead to an undersampling of the early stages of the wave propagation. These authors also showed that Moreton- and EIT-wavefronts are nearly cospatial and have a 
similar morphology, which strongly suggests that a single physical disturbance could generate both. 


A significant observational progress resulted from the next generation of EUV imagers with higher temporal cadence, such as the {\it Extreme Ultraviolet Imager} (EUVI: \opencite{Wulser04}) onboard the two spacecraft of the {\it Solar-Terrestrial Relations Observatory}
(STEREO: \opencite{Kaiser08}), the {\it Sun Watcher using Active Pixel System Detectors and Image Processing} (SWAP: \citeauthor{Halain10}, \citeyear{Halain10},
\citeyear{Halain13}; \opencite{Seaton13}) onboard the {\it Project for On-Board Autonomy 2} (PROBA2), and the {\it Atmospheric Imaging Assembly} (AIA: \opencite{Lemen12}) onboard the {\it Solar Dynamics Observatory} (SDO: \opencite{Pesnell12}). The data provided by these imaging instruments have helped to understand some long-standing problems
about the nature of the EIT waves. 
EUV waves have been observed with EUVI (\eg~ \opencite{Veronig08}; \opencite{Long11b}; \opencite{Warmuth11}) and AIA (\eg~ \opencite{Long11a}; \opencite{Liu12}; \opencite{Cheng12}).
Recently,  the simultaneous existence of more than one type of EUV wave became evident from observations taken with AIA. 
 \inlinecite{Asai12} first reported cotemporal observations of EUV and H$\alpha$ Moreton waves using SDO/AIA. They identified a dome-shaped shock front expanding outward in the corona, which would produce a sharp EUV front at low coronal heights and a Moreton wave intersecting the chromosphere. They also observed a type II radio burst consistent with the shock wavefront.

 Observations of shock-wave domes evolving outward in the corona as predicted by Uchida's model have previously been reported in a number of cases (\eg~ \opencite{Narukage04}; \opencite{Veronig10}; \opencite{Kozarev11}; \opencite{Ma11}). Along these lines, 3D MHD numerical simulations performed by \inlinecite{Selwa12} showed that twisted coronal magnetic loops can evolve into an EUV wave, which forms a dome-shaped structure that propagates in the corona after energy release in a flare followed by a dimming. The EUV wave propagates nearly isotropically on the disk and is able to produce the observed low-coronal and chromospheric signatures.

The aforementioned large-scale EUV perturbations have been historically named as EIT waves \citep[see the review by][]{Warmuth15}. In this article, to avoid referring to a specific instrument name, we use the term EUV wave in general. Moreover, we will distinguish, when needed, between their near-surface imprints in the low corona (near-surface EUV wave) and the signatures of the 3D dome seen in projection (3D-dome EUV wave). 

EUV waves are often referred to as fast and slow waves (\opencite{Liu10}; \opencite{Ma11}; \opencite{Chen11b}; \opencite{Liu14}). The fast ones are thought to be shock waves linked to chromospheric Moreton waves and coronal soft X-ray waves (\opencite{Asai12}; \opencite{White14}; \opencite{Cliver13}), while the slow ones  could be real waves or belong to the category of non-MHD traveling perturbations or pseudo waves \citep{Warmuth15}. Some authors suggest a third category of EUV waves, the hybrid ones, considering that in some events multiple bright fronts can be observed in conjunction, with some being true waves and others being pseudo waves \citep{Patsourakos12}.

 The category of pseudo waves has been proposed by several authors who proposed that EIT waves are the consequence of the magnetic field reconfiguration during a CME eruption \citep[see \eg][]{Delannee99}.
The apparent EIT wave is then the projection on the disk of an expanding CME envelope or its footpoints. Depending on the global or surrounding magnetic topology, propagating and stationary brightenings can be observed. 
\inlinecite{Delannee99} (see also \opencite{Delannee00}; \opencite{Delannee08}) proposed other non-wave model for EIT disturbances, arguing that they would result from Joule heating in electric-current shells during the opening of field lines in a CME ejection.  \citeauthor{Chen02} (\citeyear{Chen02}, \citeyear{Chen05a}, \citeyear{Chen05b}, \citeyear{Chen05c}) suggested that a propagating density enhancement near the solar surface, which appears in numerical simulations of a {rising flux rope due to the expansion of field lines}, is responsible for EIT waves; while Moreton waves correspond to the faster shock front that is a consequence of the flux-rope radial acceleration. \citeauthor{Attrill07} (\citeyear{Attrill07}, \citeyear{Attrill09}) proposed an alternative mechanism to explain 
EUV-wave diffuse fronts; they argued that they could be generated by propagating magnetic reconnections during the expansion of CME flanks.

Another long-standing problem is related to the origin of Moreton and associated coronal EUV waves and is described by the question of what drives them: flares or CMEs. It is also unclear whether it is even possible to distinguish between these two drivers.  
Since CMEs and flares are able to liberate enormous amounts of energy in a short period of time, they can both be candidates to produce strong shock-waves in the corona. During Moreton events, CMEs and flares appear to originate in the same AR and occur nearly simultaneously, so that it is difficult to discern which is primarily responsible for the coronal shock wave and the chromospheric effect. The characteristics of a shock wave depend mainly on the 3D temporal piston that generates it \citep{Vrsnak08}. Moreton waves exhibit the characteristics of a blast wave, as mentioned previously, \ie~a single shock that freely propagates and decays with time and distance. A blast wave could be generated by a 3D piston in expansion acting in a short time, as is the case of a pressure pulse caused by the flare energy deposition in the lower atmosphere. A 3D-piston driver acting for longer times and expanding in all directions, as is the case of a CME, could supply energy in a continuous
way to the shock. As a consequence, the shock would evolve faster than the piston, which would generate a supersonic wave, even in the case of a subsonic driver. In this case the shock is expected to maintain the characteristic shape of the driver.
 There are other different types of 3D pistons, \ie~a rigid body or an expanding blunt driver moving through the plasma and generating a shock cone or a hyperbole-shaped surface (called bow shock), respectively.
The capability of a piston to generate a shock wave depends mostly on its size and acceleration \citep{Temmer09}, added to the characteristics of the coronal medium in which it propagates. In this regard, a small and impulsively accelerated driver is able to generate coronal shocks.
Therefore, other phenomena are candidates for generating globally propagating coronal waves, \ie~eruptive filaments and small-scale ejecta (\opencite{Warmuth15}, and references therein). Recent multidirectional observations performed with STEREO and SDO showed a direct relation between CMEs and EUV waves (see \opencite{Patsourakos12}, and references therein).
The CME and the wave, initially cospatial, decouple after the initial stages of the CME evolution and the shock wave becomes a freely propagating MHD wave. This suggests that in its first steps the CME expansion acts as a temporary 3D  piston that drives the shock wave. In this scenario, the CME bubble should follow the coronal wave expansion.

With the aim of contributing to the understanding of EUV waves, chromospheric Moreton waves, and their sources, we present a detailed analysis of the kinematics and directional characteristics of the wave event o 29 March 2014. An X1.0 class flare (SOL2014-03-29T17:48) occurred on that date in AR 12017 (N10 W32). This event was observed by several observatories from the ground and by the {\it Interface Region Imaging Spectrograph} (IRIS: \opencite{Depontieu14}), SDO, STEREO,  the {\it Reuven Ramaty High Energy Solar Spectroscopic Imager} (RHESSI: \opencite{Lin02}), and {\it Hinode} \citep{Kosugi07}. GOES soft X-ray emission started to rise at around 17:35 UT and peaked at 17:48 UT. The flare was accompanied by a filament eruption, chromospheric and EUV waves, and a CME. The morphology of the flare onset and the magnetic field structure have been studied by \inlinecite{Kleint15} and \inlinecite{Liu15}. The event began with an asymmetric filament eruption, its western portion arched upward at $\approx$ 17:35 UT and remained quasi-static for a few minutes. This was accompanied by a sustained increase in X-ray emission observed by RHESSI. After this first stage, at $\approx$ 17:43 UT the filament started to erupt, which led to a CME. Two hard X-ray sources appeared at $\approx$ 17:45 UT within the two elongated flare ribbons seen by IRIS in 1400\,\AA. The event is front-sided from Earth's view and back-sided from the STEREO A and B spacecraft. EUV images show a global coronal bright front that begins to  expand around 17:45 UT. The coronal bright front, accompanied by a coronal dimming, expanded to the north, west and east, while H$\alpha$ high-resolution images showed a Moreton wave. These wavefronts constitute the subject matter of this article. 

Our article is organized as follows. In \sect{data} we describe the data used, \sect{analysis} presents our qualitative analysis of the wave events at different atmospheric levels (chromosphere, low corona) and of their white-light counterpart. In \sect{quantitative} we perform a quantitative study of the shock-front properties and propose a geometrical model to explain the observed wavefronts as the chromospheric and EUV 
traces of an expanding and outward-traveling bubble intersecting the Sun. Finally, in \sect{discussion} we discuss our results and conclude.

\section{The Data}
\label{sec:data}

AIA provides full-disk images of the low corona with a pixel spatial size of 0.6\arcsec and a temporal resolution of 12 sec in multiple wavelengths. This characterizes AIA as the best instrument to date for the analysis of coronal waves in the EUV range \citep[see the review by][]{Liu14}. 

To study the coronal bright fronts on 29 March 2014, we used the AIA passbands centered on Fe\,{\sc ix} (171\,\AA), Fe\,{\sc xii/xxiv} (193\,\AA) and Fe\,{\sc xiv} (211\,\AA), in which plasma emission at temperatures in the range of 0.5\,--\,2.5 MK can be detected. To allow for comparisons with the Moreton-wave data, we also analyzed images captured with the filter centered on He\,{\sc ii} (304\,\AA). The He\,{\sc ii} passband detects plasma in the transition region with a characteristic temperature of \mbox{$\approx\,5 \times 10^4$ K}.
AIA images between 17:35 UT and 18:15 UT 
were processed using the Solar Software standard procedures and were rebinned to $2048^2$ pixels to decrease memory requirements and to increase the signal-to-noise ratio. To avoid variations in the background of the images, we chose to use AIA images with exposure times longer than 0.06 s for the 171\AA, 211\AA~and 304 \AA~bands~and 0.6 s for the passband centered on 193 \AA.  This selection criterion results in time differences between consecutive images of 24 s at most. The images were derotated to a pre-event time (17:30 UT) to correct for the displacement of coronal structures due to the differential rotation of the Sun.

To study the Moreton wave, we used high-temporal resolution H$\alpha$ images obtained with the {\it H$\alpha$ Solar Telescope for Argentina} \citep[HASTA:][]{Bagala99, Fernandezborda02, Francile08}. HASTA is located at the Estaci\'on de Altura U. Cesco of the Observatorio Astron\'omico F\'elix Aguilar, in El Leoncito, San Juan, Argentina. HASTA records images at the hydrogen H$\alpha$ line center (656.27 nm,  0.03 nm FWHM) and in its red and blue wings  ($\pm$ 0.5 nm). The instrument has two operation modes: patrol mode and flare mode. In patrol mode, the camera obtains images with a cadence of 2 min. Solar activity is analyzed in real time; when an event with an intensity above a certain level is detected, the camera switches to flare mode. In this mode, HASTA can image the Sun with a five-second temporal cadence  in the H$\alpha$ line center. 
The high temporal cadence in the flare mode makes HASTA a suitable instrument to study Moreton waves; in particular, during the initial propagation phase. HASTA images have 1280 $\times$ 1024 pixels with a spatial resolution of $\approx$ 2\arcsec. In this article, we use a set of H$\alpha$ line-center images acquired with a 50-millisecond exposure time between 17:41 UT and 17:53 UT, covering the whole Moreton event with a five-second temporal cadence. The images were pre-processed following a standard instrument procedure, scaled, and rotated to match AIA frames. To correct for the jitter produced by the seeing, we applied a cross-correlation technique to center AR 12017 in all images.

\section{The Wave Event on 29 March 2014}
\label{sec:analysis}

In the following sections, we analyze propagating disturbances that evolve at different atmospheric levels using remote-sensing data recorded in several spectral bands. To measure and compare these dissimilar data, we 
defined
two measurement frameworks, a surface frame and a plane-of-sky frame. The first is intended to be used with chromosphere and transition region data, \ie~those registered in the H$\alpha$ and He {\sc ii} bands. The related emissions take place close to the solar surface, and in consequence should be measured over a curved surface, 
namely the solar sphere. The plane-of-sky frame is useful to measure emissions at low coronal heights, such as those in the AIA coronal EUV bands (\eg~193\,\AA~and 211\,\AA). The bulk of these emissions comes from a certain coronal height, but we used the data to delineate features propagating at different altitudes, like  dome-shaped shock wavefronts, whose borderline can be considered as approximately located in the plane-of-view of the analyzed images. In consequence, we define the plane-of-sky frame, which is flat as opposed to the surface frame.

 \begin{figure}    
   \centerline{
							 \includegraphics[width=1.0\textwidth,clip=]{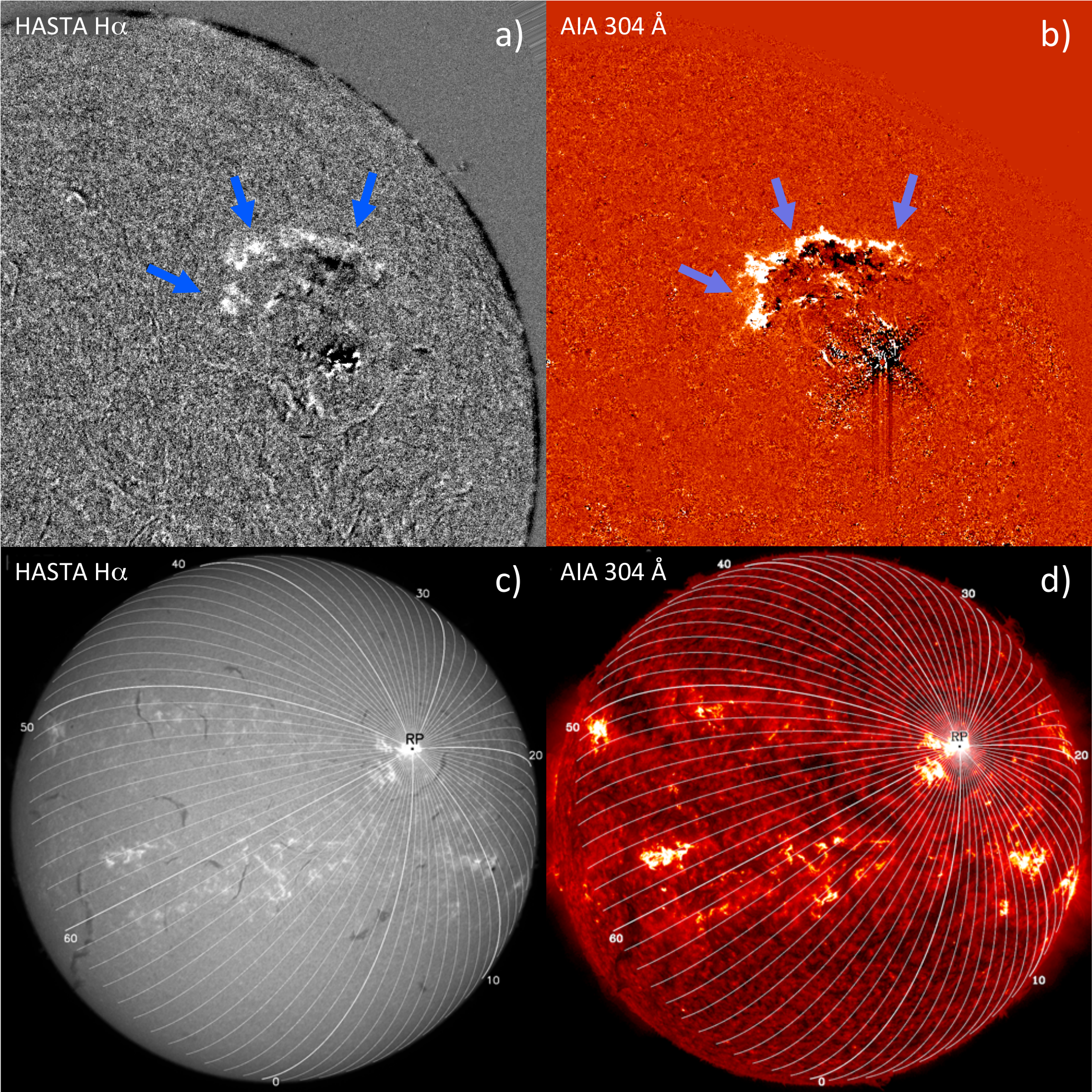}
              }
              \caption{(a) A view of the Moreton wave in a  running-difference H$\alpha$ image at 17:49:42 UT (see \href{figures:./20140329-Halpha-rd.mov}{Movie-Halpha}). (b) Similar to panel a in He\,{\sc ii} at 17:49:31 UT  (see \href{figures:./20140329-304-rd.mov}{Movie-304}). The light blue arrows in both panels point to the propagating perturbation. (c) Great circles traced on an image recorded at the center of the H$\alpha$ line at 17:43:53 UT. (d) Similar to panel c in He\,{\sc ii} band at 17:38:07 UT. The circles are traced on the surface of the solar sphere, \ie~with a radius of 1\,\Rsun, and depart from the radiant point defined as the site where the flare is strongest (coordinates N10.3 W32.8). The 72 sectors separated by 5$\degree$ each are arbitrarily numbered in a counterclockwise order, beginning from the sector that is centered on the solar south pole. Some of the sectors have been numbered. The tracing corresponds to an angular span of $\alpha=$100$\degree$ from the radiant point. The field of view in panels c and d is 2.08$\times$2.08\,\Rsun. }
   \label{fig:CirclesSurf}
   \end{figure}

\subsection{H$\alpha$ and He\,{\sc ii} Wave}
\label{sec:chromosphere}

The Moreton-wave event on 29 March is detectable in HASTA images applying running-difference techniques.  It appears as a diffuse arc-shaped front propagating to the north of AR 12017 and evolving after the flare impulsive phase. Similar features can be detected in the AIA He\,{\sc ii} (304\,\AA) band. A running-difference full-disk image movie built from H$\alpha$ data is attached to this article (see \href{figures:./20140329-Halpha-rd.mov}{Movie-Halpha}), as well as a running-difference subfield image movie in the He\,{\sc ii} (304\,\AA) band (see \href{figures:./20140329-304-rd.mov}{Movie-304}).

We determined the  kinematic characteristics of the Moreton wave using the intensity profiles technique \citep{Vrsnak02}. The intensity profiles [$\delta({\bf r}_i, t_j)$] are obtained from the running-difference image corresponding to time $t_j$ along paths ${\bf r}_i$ traced on the solar disk (surface frame). The paths ${\bf r}_i$  are the plane projection of great circles of the solar sphere passing through a radiant point (RP, see \fig{CirclesSurf}c and \ref{fig:CirclesSurf}d). We approximated the actual origin site of the wave event, \ie~the RP,  as the centroid of the most intense flare kernel in the AR that corresponds to 514.2\arcsec~in the east--west  and 263.4\arcsec~ in the north--south directions in the heliocentric-Cartesian coordinate system and N10.3 W32.8 in the heliographic system.

The intensity profiles [$\delta({\bf r}_i, t_j)$] were discretized in angular steps along the corresponding great circle with a resolution of $0.1\degree$, starting from the RP. Each discrete value is computed  by averaging  the intensity values laterally over an angular sector of 5$\degree$ centered on the RP. In this way, the solar surface distance from the RP to a certain point $P$ of the path ${\bf r}_i$ can be computed as $d_c=\alpha$ 1\,\Rsun, where $\alpha$ is the angle subtended between the two radial vectors formed by the pairs of points $(O,RP)$, $(O,P)$ in 3D, where $O$ is the center of the solar sphere.

To investigate the angular dependence of the evolution of the wave event, we applied a stack plot procedure \citep{Liu10, Li12}. In this way, distance-time (DT) maps werebuilt by stacking in columns the intensity profiles [$\delta({\bf r}_i, t_j)$] corresponding to a specific path ${\bf r}_i$ over the full time-span of the wave event. These columns were expanded 
for the time span in seconds between the corresponding image of the set and the next one, therefore the DT-maps abscissae  are in
seconds. The DT-map ordinates are the surface distance [$d_c$] measured from the RP.

  \begin{figure}    
   \centerline{\includegraphics[width=1.0\textwidth,clip=]{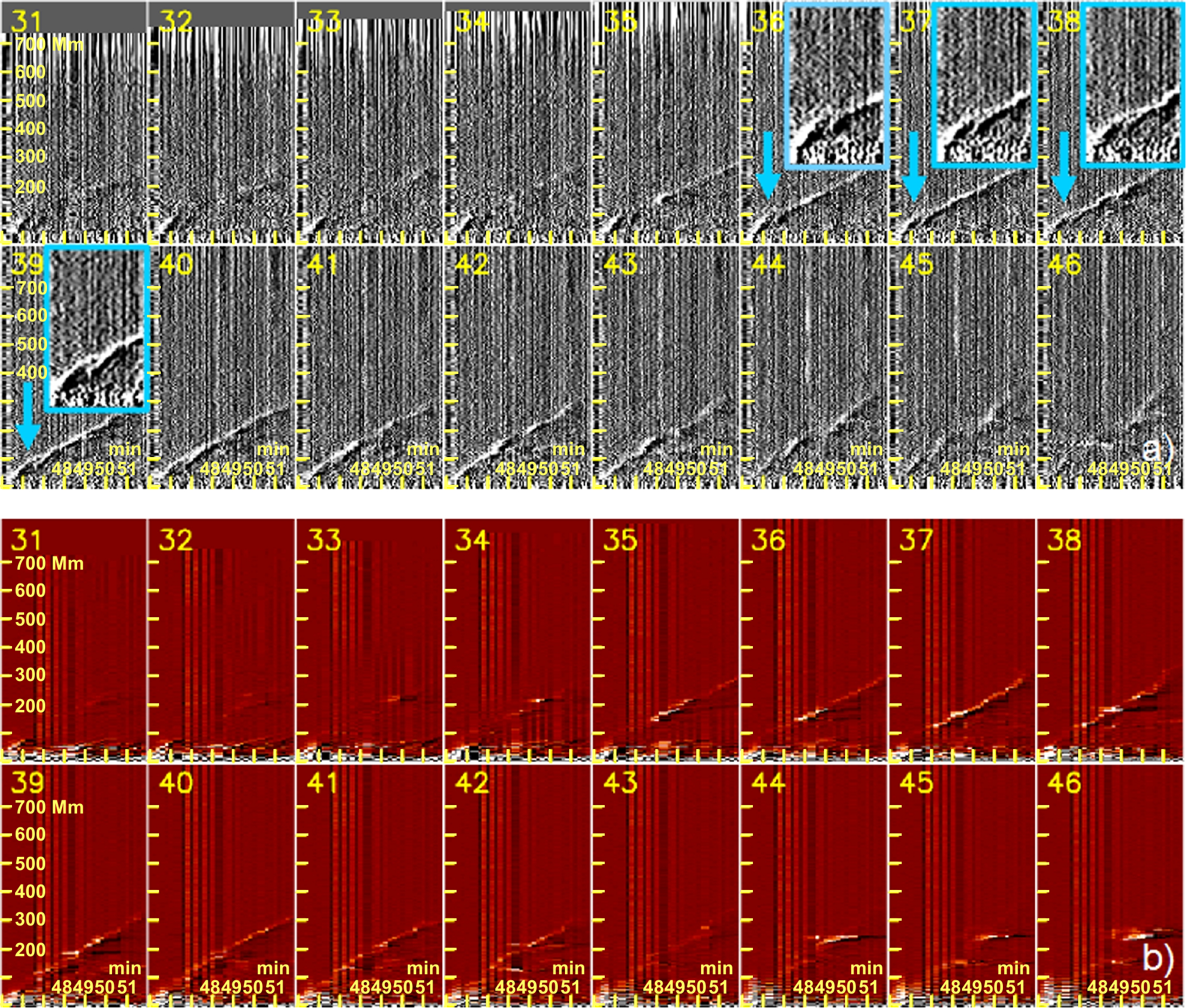}}
\caption{Stack plots of sectors 31  to 46. (a) Corresponds to H$\alpha$. The insets in the panels for sectors 36 to 39 show enlargements of the portion of the figure pointed out by a light blue arrow. (b) Corresponds to He {\sc ii}. The abscissae represent the time in seconds and the ordinates the distance [$d_c$] from the RP. The field of view is 850\,Mm $\times$ 420 sec. The time ranges from 17:45:00 UT to 17:52:00 UT.}
   \label{fig:StackSurf}
   \end{figure}

We covered the 360$\degree$ around the RP with 72 sectors of 5$\degree$ each, thus obtaining 72 DT maps in correspondence with the 72 paths ${\bf r}_i$, numbered counterclockwise starting from a great circle passing through the solar south pole. \fig{CirclesSurf} shows the 72 sectors superimposed to H$\alpha$ (panel c) and He\,{\sc ii} (panel d) images. The reference sector $0$ starts at the great circle denoted with the thick white-line that crosses the solar south pole.

The results of this procedure for sectors 31 to 46 can be observed in \fig{StackSurf}. The Moreton wave is visible in H$\alpha$ DT-maps (\fig{StackSurf}a) as an oblique bright trace for all the displayed sectors. The slope of these traces indicates the speed of the Moreton wave in the corresponding sector. The region where the traces are brighter, \ie~between sectors 35 to 46, shows an initial brief lapse of deceleration after which the speed of the Moreton wave is almost constant. 
Sectors 36 to 39 show regions with an apparent overlap of two traces very close in time (see the insets in the corresponding sector panels).
He {\sc ii} DT-maps (\fig{StackSurf}b) show similar oblique traces. 
The traces are not continuous and have the appearance of a succession of parcels with fluctuating intensity. No overlap of the traces can be discerned, in contrast to the H$\alpha$ DT-maps. The vertical stripes in the background correspond to changes in the exposure time of AIA during the recording of the flare peak-intensity.

\begin{figure}    
  \centerline{
							\includegraphics[width=1.0\textwidth,clip=]{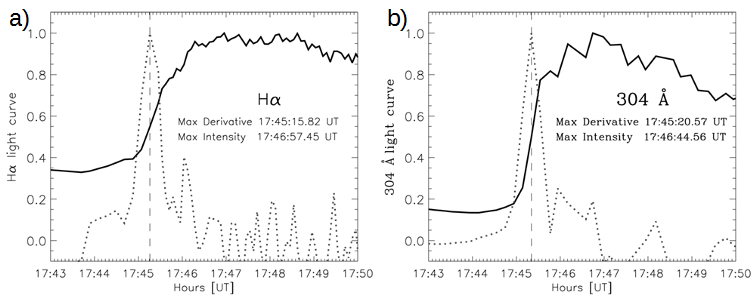}
						 }
						  \caption{Flare intensity (solid line) and its derivative (dotted line) for: (a) H$\alpha$ and (b) He {\sc ii} images. The curves are normalized to their maximum intensity. The peak times of the derivative curves are indicated with a vertical dashed line. We choose the peak time of the H$\alpha$ derivative curve,  t = 17:45:15.82 UT 
							as the reference flare onset time [t$_{on}$ = 17:45:16 UT]. }
   \label{fig:FlareInt}
   \end{figure}

We obtained normalized light curves of the whole flaring region, shown in \fig{FlareInt} for H$\alpha$ (panel a) and He {\sc ii} (panel b). To compare these curves, which have a similar temporal evolution but a different intensity scale, we divided them by their maximum intensity value, {\it i.e.} we normalize them. From these curves we derived a flare onset time as the peak of the derivative of the light curves. The
derivative was computed using a standard quadratic three-point Lagrangian interpolation. The values obtained are t = 17:45:15.82 UT  
for H$\alpha$ and t = 17:45:20.57 UT
for He {\sc ii}. We chose the value obtained from H$\alpha$ as the reference flare onset time [t$_{on}$ = 17:45:16] because of the higher temporal cadence of the HASTA telescope in flare mode.

  \begin{figure}    
   \centerline{
							 \includegraphics[width=1.0\textwidth,clip=]{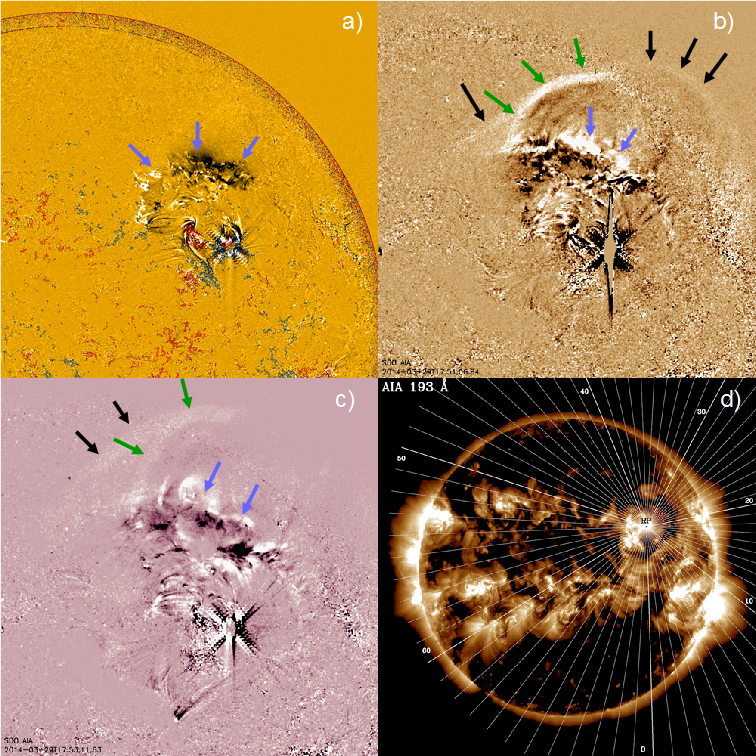}
              }
              \caption{Snapshots of the event in the low corona. (a) Running-difference image from AIA 171\,\AA~at 17:50:23 UT, with HMI contours $\pm$100 and $\pm$500 G superimposed (blue/red corresponds to positive/negative values) (see \href{figures:./20140329-171-rd.mpg}{Movie-171}). (b) Running-difference image from AIA 193\,\AA~at 17:51:06 UT (see \href{figures:./20140329-193-rd.mpg}{Movie-193}). (c) Running-difference image from AIA 211\,\AA~at 17:53:11 UT (see \href{figures:./20140329-211-rd.mpg}{Movie-211}). (d)
 Full-disk direct image from AIA 193\,\AA~with sectors traced under a plane-of-sky assumption. We have numbered some of the 72 defined sectors. The black arrows in panels b and c indicate the faster shock, the green arrows a slower shock-like propagating feature, and the light blue arrows a propagating perturbation closer to the solar surface (see also panel a).}
   \label{fig:LowCorona}
   \end{figure}

\subsection{EUV waves}
\label{sec:low-wave}

Several propagating features with dissimilar characteristics are visible in AIA images at 171\,\AA, 193\,\AA, and 211\,\AA, some of which are displayed in panels a, b, and c of \fig{LowCorona}. Movies of running-difference images in these bands accompany this article (see \href{figures:./20140329-171-rd.mpg}{Movie-171}, \href{figures:./20140329-193-rd.mpg}{Movie-193}, and \href{figures:./20140329-211-rd.mpg}{Movie-211}). \fig{LowCorona}a shows a running-difference 171\,\AA~image at t\,=\,17:50:23 UT superimposed on magnetic field contours of a {\it Helioseismic Magnetic Imager} (HMI: \opencite{Scherrer12}) magnetogram. The contours show the complex magnetic structure of AR 12017, while the large-scale loops to the east indicate its connectivity to scattered negative-polarities belonging to a large bipolar region. The light blue arrows in this panel indicate a northward-propagating arc-like disturbance that we associate with the chromospheric Moreton-wave event described in Section~\ref{sec:chromosphere}. 

\fig{LowCorona}b and \ref{fig:LowCorona}c exhibit a tenuous circular-shaped propagating feature denoted by black arrows. The running-difference 193\,\AA~and 211\,\AA~images appear similar, with this feature surpassing the solar disk. This structure can be associated with an MHD coronal wave- or shock-front that likewise propagates in all directions. A second slower and circular feature indicated with green arrows in the same panels is also noticeable. This might be 
a secondary shock, but it could also be attributed to the leading edge of the expanding CME, as suggested by \citet{Patsourakos12}.
Light blue arrows in \fig{LowCorona}b and \ref{fig:LowCorona}c indicate more irregular propagating features that show correspondence with these fronts, also pointed out with light blue arrows, in \fig{LowCorona}a. They could be ascribed to an origin closer to the solar surface. They are probably caused by the interaction of the shock fronts at transition region levels or by effects of the lateral expansion of the CME bubble during its ejection. The analysis of the kinematics of the various fronts in \sect{quantitative} is helpful to validate some of these hypotheses.

To follow and characterize the visible fronts in the plane of view, we now track the evolution of the wavefronts in the plane-of-sky frame. To accomplish this, we built a new set of stack plots obtained from a plane-of-sky measurement scheme, obtaining the intensity profiles in a similar way as described in \sect{chromosphere}, \ie~discretizing in linear steps of $\sim$ 1215 km along the corresponding straight path that bisects the sectors, starting from the RP. Each discrete value was computed by  averaging  the intensity values laterally over an angular sector of 5$\degree$ centered on the RP.
To ensure that the surface and plane-of-sky frames coincide, we traced the plane-of-sky bisector for each sector
to match the solar-surface great circle tracing at some points P distant $\alpha$\,=\,1$\degree$ from the RP, where $\alpha$ is the angle subtended between the two radial vectors formed by the pairs of points $(O,RP)$ and $(O,P)$, as defined above. The resulting sector tracing of the plane-of-sky frame can be observed in \fig{LowCorona}d. 

The plane-of-sky DT-maps obtained by stacking the intensity profiles of the 193\,\AA~band for various sectors can be observed in \fig{Stack193}. Each panel of the mosaic is the DT map of a specific sector, but in this case the DT-map ordinates exceed the solar limb, which is indicated by a horizontal white line in the different panels. Although the wave event is visible in sectors 10 to 47, we only show it from sector 16 onward in this figure. The wave feature appears brighter from sector 35 to 44, where it has an angular span of approximately  45$^\circ$ toward the north. \fig{Stack193} shows the following:

 \begin{figure}    
   \centerline{
							 \includegraphics[width=1.0\textwidth,clip=]{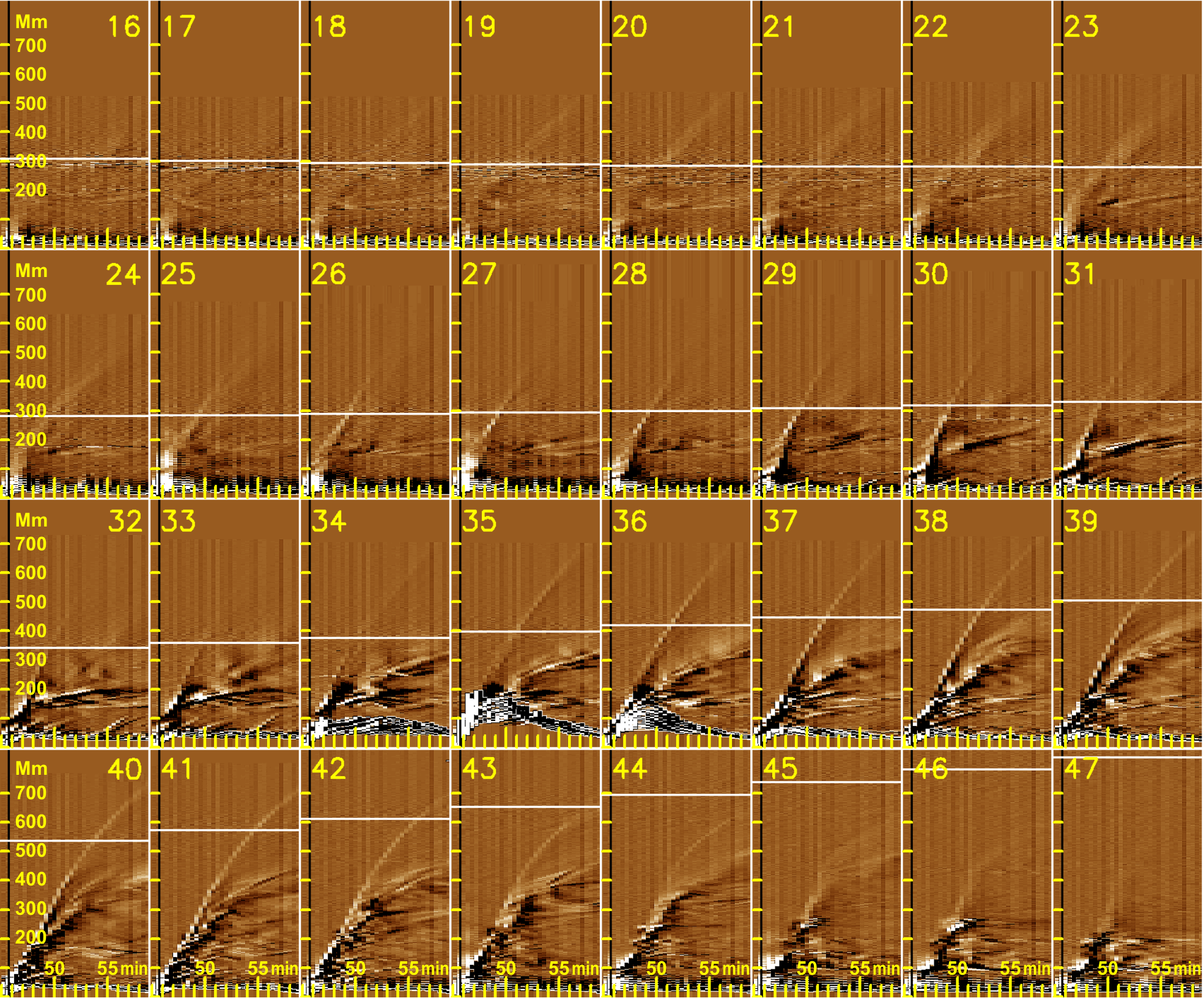}
              }
   \caption{Stack plots of sectors 16 to 47 from AIA 193\,\AA~in the plane-of-sky frame. The abscissae represent time in seconds and the ordinates distance [$d_s$] from the RP in Mm. The dimensions of each panel are 850.3 Mm $\times$ 840 sec (t\,=\,17:45:00 UT to t\,=\,17:59:00 UT). The vertical black line indicates the reference flare time. The horizontal white line in each stack plot indicates the solar-limb position.  The separation between two ticks in the abscissae is 60 seconds and in the ordinates 100 Mm.}
   \label{fig:Stack193}
   \end{figure}

\begin{itemize}
\item{The first visible perturbation is a well-defined bright front with an initial rising slope nearly constant, which tends to decelerate far from the RP. The front surpasses the limb indicating 
it 
propagates out of the Sun. This well-defined thin front can be associated to a shock that compresses the coronal medium as it propagates producing an EUV emission  enhancement. Close to the RP it is hard to discern the front due to the flare intensity effects that disturb the observation. This shock front would correspond to the borderline of a 3D dome-shaped structure,  evolving outward in the corona. }
\item{The slopes of the first front are similar but not equal in the different panels, suggesting an angular dependence of the shock-front speed. This could be attributed to inhomogeneities of the coronal environment where the wave propagates, 
but a line-of-sight projection of the 3D dome-shaped structure or a particular behavior of the 3D piston that generates the shock wave are also possible.} 
\item{After the shock, some bright and complex propagating structures are visible in all of the displayed sectors. It is possible to see that these features, initially co-spatial, detach from the shock front at a certain distance from the RP (see panels for sectors 35 to 45). This can be attributed to near-surface EUV waves.}
\item{Some diffuse features, slower than the shock but faster than the bright structures, surpass the solar limb in panels for sectors 30 to 32, which may be indicative of an outward propagation. Probably, these features correspond to some CME parts visible during its lift off, but the presence of other true MHD waves cannot be ruled out.}
\end{itemize}

 \begin{figure}    
   \centerline{
							 \includegraphics[width=1.01\textwidth,clip=]{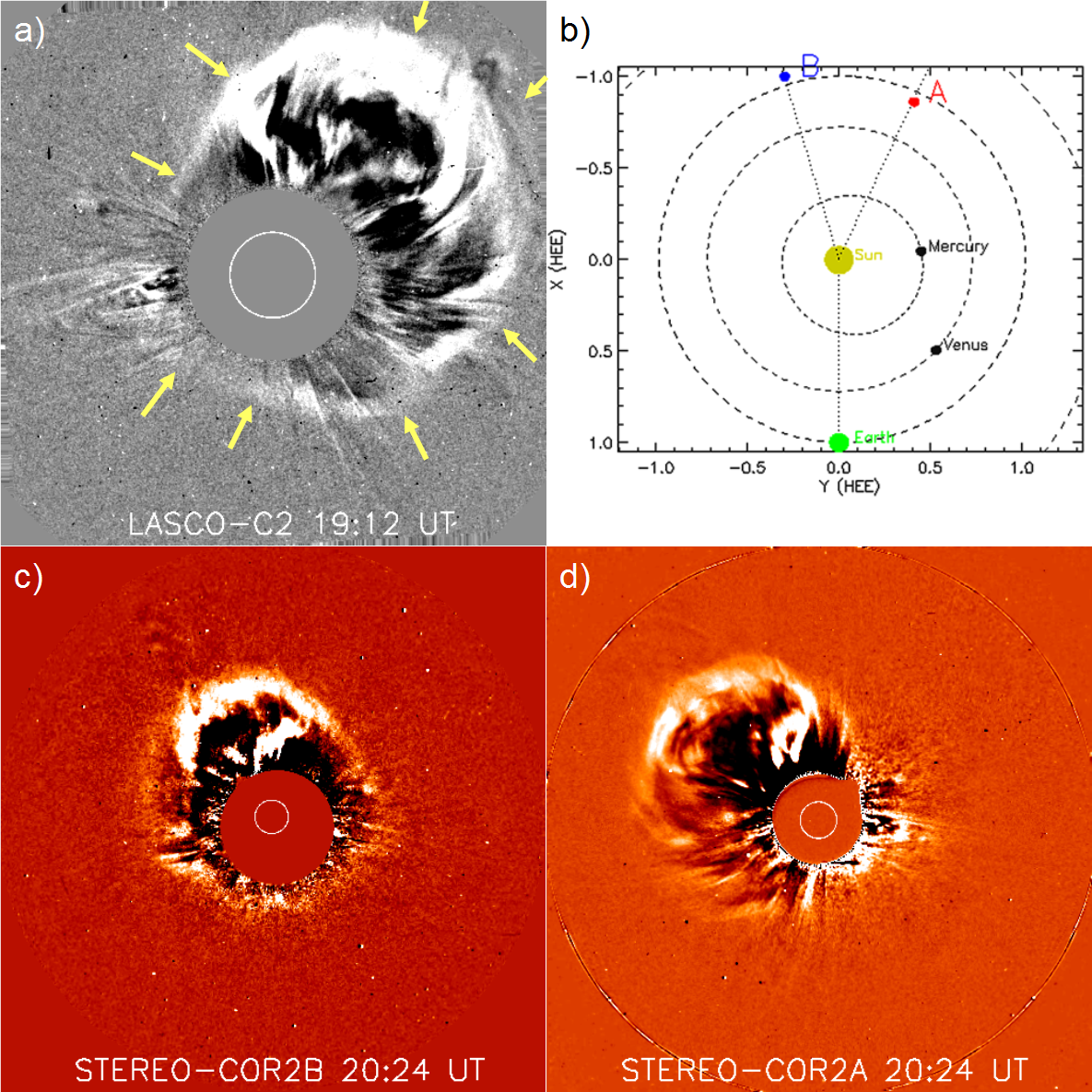}
              }
              \caption{The associated CME as viewed by the coronagraphs LASCO from Earth's perspective and the two COR2 from the STEREO spacecraft vantage points. (a) LASCO C2 running-difference image with arrows indicating the shock driven by the CME. (b) Relative locations of the spacecraft as shown by the tool ``Where is STEREO'' at the STEREO Science Center (\url{http://stereo-ssc.nasa.gov}). (c) Running-difference image from COR2 on STEREO-B. (d) Running-difference image from COR2 on STEREO-A.}
   \label{fig:CME}
   \end{figure}

\subsection{White-light Counterpart}
\label{sec:corona-WL}

The CME associated with the low-coronal event was first seen at 18:12 UT with the C2 coronagraph of the {\it Large Angle and Spectroscopic Coronagraph} (LASCO: \opencite{Brueckner95}) onboard SOHO. Initially, it appeared as a bright front toward the north--west. The CME quickly evolved to become a full-halo CME with its main bulk also to the north--west (see \fig{CME}a). According to the SOHO LASCO CME Catalog (\url{http://cdaw.gsfc.nasa.gov/CME_list/}), the projected speed of its fastest front at a position angle of 324$\degree$ after a linear fit is 528 km s$^{-1}$, with the most distant points showing a slightly decelerated profile. The STEREO spacecraft were located almost at the opposite side of Earth, separated by almost 42$\degree$, as shown in the cartoon in \fig{CME}b. From the viewpoint of the coronagraph COR2 onboard STEREO-B, the CME appears nearly symmetric (\fig{CME}c) but with its main bulk towards the north. From the perspective of STEREO-A, the CME appears as a partial halo traveling toward the northwest (\fig{CME}d). With these stereoscopic considerations, the 3D propagation direction of the CME is obtained by means of the forward model developed by \citet{Thernisien09}. The deduced direction is N37 W34, suggesting a northward deflection of $\approx27\degree$ when we consider the source region at the RP coordinates (N10.3 W32.8). 

We interpret the outermost rim of the CME as indicative of a shock wave ahead of the CME bulk, as done by \citet{Ontiveros09}. To enhance the shock signatures, which are fainter than the bulk of the CME, we produced sequences of running-difference images. The CME shock can be discerned from the much brighter CME material, not only because it is fainter, but also because it appears as a distinct somewhat circular and continuous rim that envelops the CME bulk, which is usually of-center because of projection effects.
Toward the northwest in \fig{CME}a the CME leading edge meets the shock, while especially to the east and south the only observable feature is the shock, as indicated by the arrows. In a similar manner, in the STEREO-B COR2 image (\fig{CME}c) the bulk of the CME travels toward the north, while to the west, south, and east only the shock is evident. STEREO-A COR2 also shows the shock wave but to the southeast, in agreement with the CME propagation direction and the spacecraft locations. An interpretation of these images to find correspondences between the CME and solar features is attempted in \sect{link-WL}.

\section{The Kinematics of the Moreton and EUV Waves}
\label{sec:quantitative}

\begin{figure}    
	\centerline{
							\includegraphics[width=1.0\textwidth,clip=]{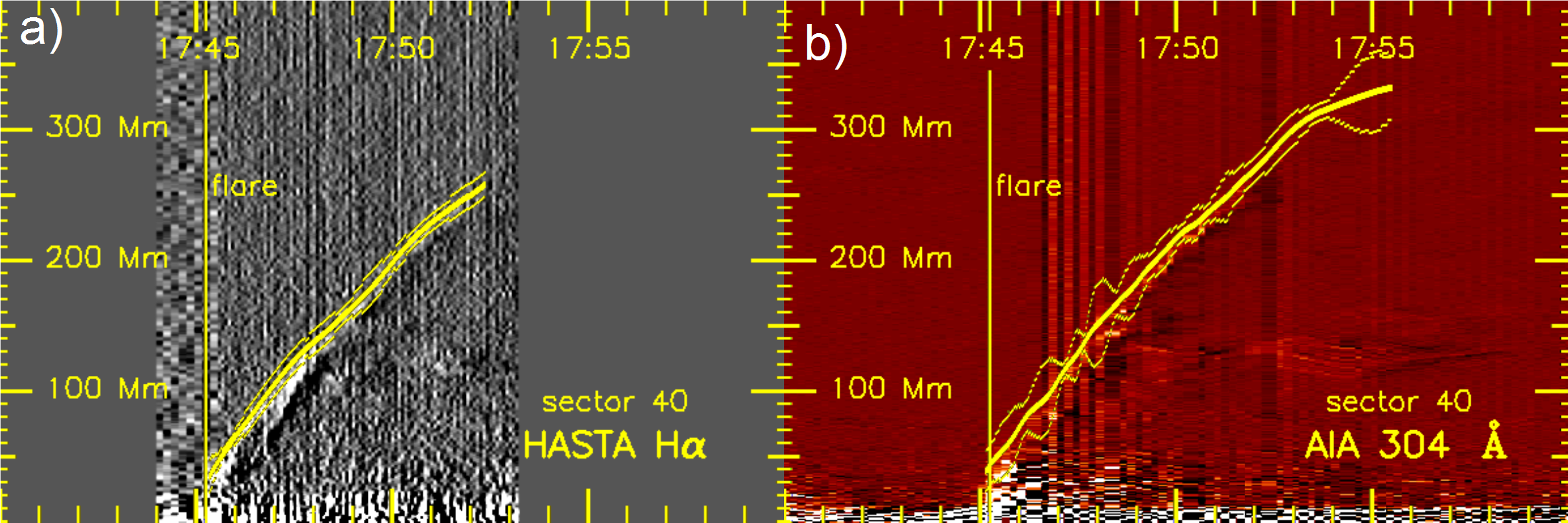}
	           }						
             \caption{Sample DT map for sector 40 ($d_c$ {\it vs.} time). (a) Corresponds to HASTA data. (b) Corresponds to AIA 304\,\AA~ data. The average spline is drawn with a thick yellow line and outlines the fastest front. The thin yellow lines at both sides mark an error band at 3$\sigma$s. The vertical yellow line indicates the flare onset-time.}
   \label{fig:Spline}
   \end{figure}

To measure the locations of the wavefronts in the DT maps obtained from the surface-frame and the plane-of-sky assumptions, we applied a visual method to identify and select the bright traces given that the fronts exhibit a non-homogeneous and diffuse pattern,  especially away from the RP. This task was performed on each DT map by manually selecting several points belonging to the leftmost side of the bright trace outline, \ie~the features of the traveling perturbation that appear first in time. These points were then connected by means of a spline to build a profile that represents the temporal evolution of the distance [$d_c$]. To minimize errors the procedure was repeated at least five times, obtaining an average spline for each of the sectors of interest.
As an example, \fig{Spline} shows the average splines obtained under the surface-frame assumption for sector 40 for H$\alpha$ (panel a) and He {\sc ii} (panel b) superimposed on the DT map. The thin yellow lines at both sides of the thick line mark the error band at $\pm$ 3$\sigma$s.

CCCCCCCCCCCCCCCCCCCC

\subsection{Surface Velocity}
\label{sec:velocity-disk}

 We estimate the kinematic characteristics of the Moreton wavefronts by fitting power-law curves  \citep{Warmuth04a, Francile13} for every sector of interest to the average spline obtained, as indicated above. These curves are suitable to fit smoothly kinematic trajectories with non-constant accelerations. The power-law is given by: 

   \begin{equation}  \label{eq:power}
     d(t) = c_1 (t-t_0)^{\delta} + c_2,
   \end{equation}

\noindent	
where $t_0$ is given the arbitrary value of 17:30:00 UT. To understand the kinematic evolution, 
the instantaneous values of speed and acceleration are evaluated at the flare onset time t$_{on}$ = 17:45:16 UT	and at a subsequent time t = 17:50:00 UT using the following equations:

   \begin{equation}  \label{eq:vel}
     v(t) = c_1 \ \delta \ (t-t_0)^{(\delta - 1)},
   \end{equation}

   \begin{equation}  \label{eq:acel}
     a(t) = c_1 \ \delta \ (\delta-1) \ (t-t_0)^{(\delta - 2)},
   \end{equation}

The instantaneous values obtained from both  H$\alpha$ and He\,{\sc ii} data are listed in Table~\ref{tableHa} and  Table~\ref{table304}, respectively. The mean values listed in the tables are obtained averaging those sectors in which the wave is brighter, \ie~sectors 35 to 43.

 In both tables the accelerations are negative with an absolute value that decreases with time, thus indicating that the wave tends to slow down with a non-constant deceleration. The absolute values obtained from He\,{\sc ii} data are on average similar to those from H$\alpha$, both in regard to acceleration and speed. The initial speeds of the Moreton front derived from H$\alpha$ at t$_{on}$ (Table~\ref{tableHa}) vary in the range 650\,--\,900 \kms, disregarding the discordant value of sector 33. The wave speed decays to approximately 490\,--\,800 \kms~ five minutes later.

To validate these measurements, we applied the automated Coronal Pulse Identification and Tracking Algorithm (CorPITA: \opencite{Long14}) to HASTA H$\alpha$ data. The average values are listed in the last row of Table~\ref{tableHa}; a similar value of average initial speed but an almost double value of initial deceleration are shown. This inconsistency could be due to the method used by the CorPITA code to track the center of the perturbation profiles by means of a Gaussian fit, which detects more accurately the instantaneous values of deceleration.

\begin{table}
\caption{Accelerations and speeds of the wavefront derived from  H$\alpha$ observations  using a power-law fit. The last row displays the average acceleration and velocity obtained using CorPITA.}
\label{tableHa}
\begin{tabular}{ccccc} 
\hline
H$\alpha$  & \multicolumn{2}{c}{acceleration [\kmss]}& \multicolumn{2}{c}{velocity [\kms]} \\
 Sector      &  t$_{on}$=17:45:16 UT & t=17:50:00 UT &  t$_{on}$=17:45:16 UT & t=17:50:00 UT  \\
\hline
31  &  -0.343 $\pm$ 0.008  &  -0.233 $\pm$ 0.006  &  714.6 $\pm$ ~19.5  &  634.6 $\pm$ ~17.6 \\
32  &  -0.341 $\pm$ 0.004  &  -0.229 $\pm$ 0.003  &  653.7 $\pm$ ~~9.5  &  574.5 $\pm$ ~~8.5 \\
33  &  -0.284 $\pm$ 0.003  &  -0.187 $\pm$ 0.002  &  488.5 $\pm$ ~~6.0  &  423.1 $\pm$ ~~5.3 \\
34  &  -0.363 $\pm$ 0.005  &  -0.236 $\pm$ 0.004  &  571.1 $\pm$ ~~9.5  &  488.1 $\pm$ ~~8.2 \\
35  &  -0.321 $\pm$ 0.006  &  -0.217 $\pm$ 0.004  &  655.0 $\pm$ ~14.8  &  580.2 $\pm$ ~13.3 \\
36  &  -0.336 $\pm$ 0.006  &  -0.228 $\pm$ 0.004  &  701.1 $\pm$ ~15.2  &  622.6 $\pm$ ~13.8 \\
37  &  -0.354 $\pm$ 0.006  &  -0.241 $\pm$ 0.004  &  753.5 $\pm$ ~14.8  &  670.7 $\pm$ ~13.4 \\
38  &  -0.352 $\pm$ 0.007  &  -0.238 $\pm$ 0.005  &  726.4 $\pm$ ~16.2  &  644.2 $\pm$ ~14.6 \\
39  &  -0.405 $\pm$ 0.007  &  -0.272 $\pm$ 0.005  &  785.6 $\pm$ ~15.7  &  691.5 $\pm$ ~14.0 \\
40  &  -0.356 $\pm$ 0.006  &  -0.239 $\pm$ 0.004  &  694.5 $\pm$ ~13.2  &  611.7 $\pm$ ~11.9 \\
41  &  -0.371 $\pm$ 0.007  &  -0.249 $\pm$ 0.005  &  705.5 $\pm$ ~15.6  &  619.4 $\pm$ ~13.9 \\
42  &  -0.405 $\pm$ 0.006  &  -0.268 $\pm$ 0.004  &  708.9 $\pm$ ~12.2  &  615.3 $\pm$ ~10.7 \\
43  &  -0.387 $\pm$ 0.004  &  -0.260 $\pm$ 0.003  &  742.3 $\pm$ ~~8.7  &  652.5 $\pm$ ~~7.8 \\
44  &  -0.413 $\pm$ 0.012  &  -0.281 $\pm$ 0.008  &  896.1 $\pm$ ~29.8  &  799.6 $\pm$ ~27.0 \\
45  &  -0.337 $\pm$ 0.010  &  -0.234 $\pm$ 0.007  &  860.8 $\pm$ ~30.2  &  781.2 $\pm$ ~27.9 \\
46  &  -0.409 $\pm$ 0.017  &  -0.279 $\pm$ 0.012  &  894.2 $\pm$ ~44.4  &  798.5 $\pm$ ~40.3 \\
mean&  -0.365 $\pm$ 0.029  &  -0.246 $\pm$ 0.018  &  719.2 $\pm$ ~37.9  &  634.2 $\pm$ ~33.9 \\
\hline
CorPITA  & \multicolumn{2}{c}{-0.68   $\pm$ 0.21 [\kmss] }& \multicolumn{2}{c}{712.3 $\pm$ 58.6 [\kms]} \\
\hline
\end{tabular}
\end{table}

\begin{table}
\caption{Accelerations and speeds derived from 304\,\AA~observations using a power-law fit.}
\label{table304}
\begin{tabular}{ccccc} 
  \hline
Sect  & \multicolumn{2}{c}{acceleration [\kmss]}& \multicolumn{2}{c}{velocity [\kms]} \\
	 304\,\AA      &  t$_{on}$=17:45:16 UT & t=17:50:00 UT & t$_{on}$=17:45:16 UT & t=17:50:00 UT  \\
\hline
31  &  -0.248 $\pm$ 0.008  &  -0.166 $\pm$ 0.005  &  461.2 $\pm$ ~16.5  &  403.7 $\pm$ ~14.7 \\
32  &  -0.297 $\pm$ 0.006  &  -0.199 $\pm$ 0.004  &  569.6 $\pm$ ~12.2  &  500.7 $\pm$ ~10.9 \\
33  &  -0.305 $\pm$ 0.011  &  -0.207 $\pm$ 0.008  &  633.2 $\pm$ ~26.9  &  562.1 $\pm$ ~24.3 \\
34  &  -0.298 $\pm$ 0.009  &  -0.201 $\pm$ 0.006  &  602.9 $\pm$ ~19.9  &  533.5 $\pm$ ~17.9 \\
35  &  -0.292 $\pm$ 0.004  &  -0.197 $\pm$ 0.003  &  584.1 $\pm$ ~~9.3  &  516.2 $\pm$ ~~8.3 \\
36  &  -0.314 $\pm$ 0.004  &  -0.212 $\pm$ 0.002  &  649.7 $\pm$ ~~8.7  &  576.5 $\pm$ ~~7.8 \\
37  &  -0.304 $\pm$ 0.006  &  -0.207 $\pm$ 0.004  &  640.1 $\pm$ ~15.1  &  569.0 $\pm$ ~13.6 \\
38  &  -0.363 $\pm$ 0.006  &  -0.247 $\pm$ 0.004  &  779.0 $\pm$ ~14.6  &  694.1 $\pm$ ~13.2 \\
39  &  -0.354 $\pm$ 0.005  &  -0.241 $\pm$ 0.003  &  779.2 $\pm$ ~11.9  &  696.4 $\pm$ ~10.8 \\
40  &  -0.399 $\pm$ 0.005  &  -0.265 $\pm$ 0.003  &  722.1 $\pm$ ~~9.3  &  629.9 $\pm$ ~~8.2 \\
41  &  -0.350 $\pm$ 0.005  &  -0.237 $\pm$ 0.004  &  724.0 $\pm$ ~12.5  &  642.3 $\pm$ ~11.3 \\
42  &  -0.345 $\pm$ 0.005  &  -0.234 $\pm$ 0.003  &  715.1 $\pm$ ~10.9  &  634.7 $\pm$ ~~9.8 \\
43  &  -0.388 $\pm$ 0.005  &  -0.264 $\pm$ 0.004  &  823.5 $\pm$ ~13.2  &  732.8 $\pm$ ~11.9 \\
44  &  -0.396 $\pm$ 0.006  &  -0.269 $\pm$ 0.004  &  850.2 $\pm$ ~14.3  &  757.6 $\pm$ ~13.0 \\
45  &  -0.422 $\pm$ 0.007  &  -0.288 $\pm$ 0.005  &  925.6 $\pm$ ~19.1  &  826.8 $\pm$ ~17.4 \\
46  &  -0.418 $\pm$ 0.003  &  -0.285 $\pm$ 0.002  &  909.8 $\pm$ ~~8.2  &  812.0 $\pm$ ~~7.4 \\
mean & -0.345 $\pm$ 0.036  &  -0.234 $\pm$ 0.024  &  713.0 $\pm$ ~76.7  &  632.4 $\pm$ ~69.6 \\
\hline
\end{tabular}
\end{table}

\begin{figure}    
   \centerline{
	             \includegraphics[width=1.0\textwidth,clip=]{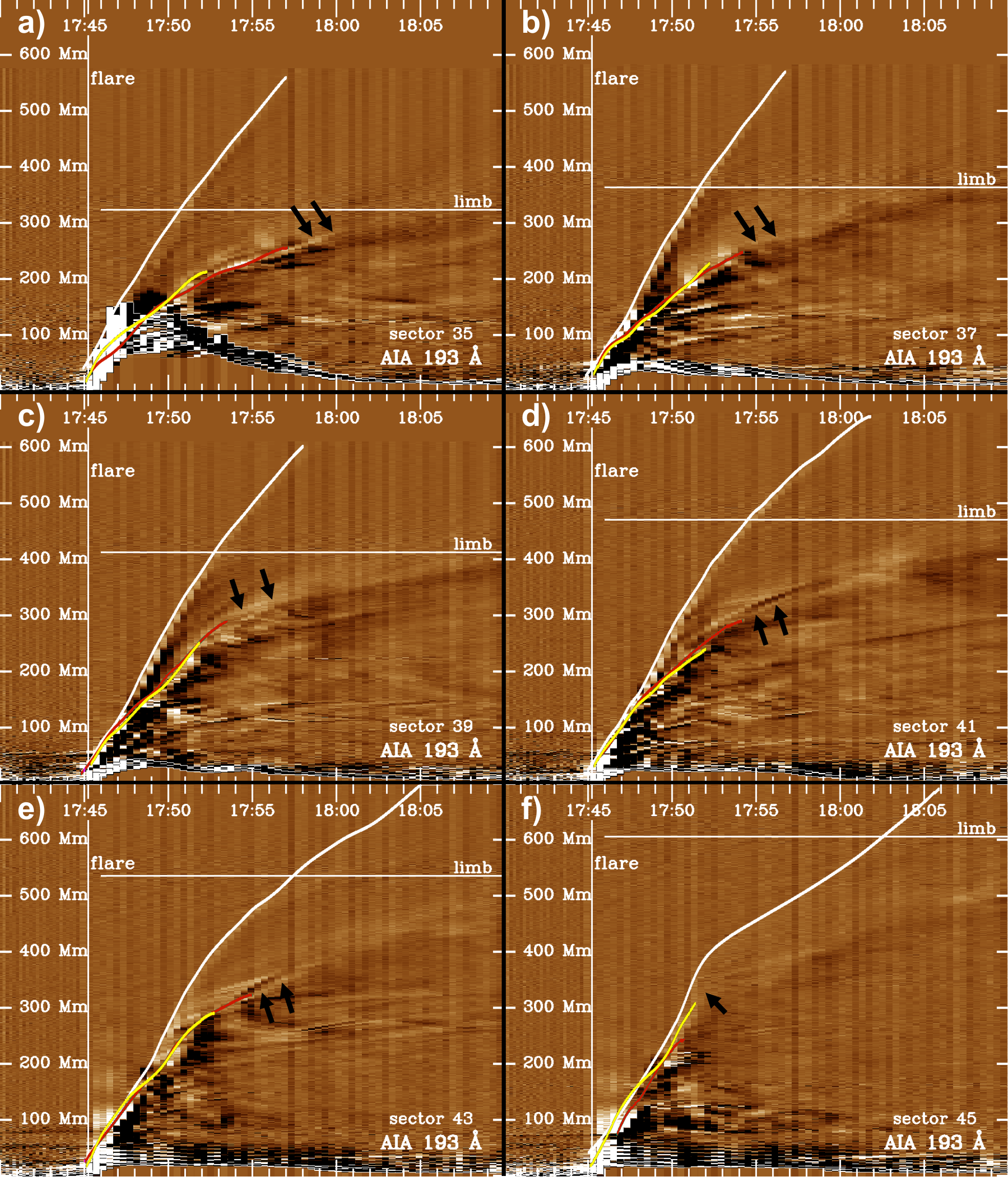}
						  }
              \caption{DT maps determined from AIA 193\,\AA~data with superimposed splines in the plane-of-sky frame outlining the temporal evolution ($d_s$ {\it vs.} time) of various features, namely: the fastest front detected with AIA 193\,\AA~(white line), the Moreton-wavefront recorded by HASTA (yellow line), and the wavefront detected by AIA 304\,\AA~(red line). The black arrows denote bright coronal features that appear to persist after the traces derived from chromospheric data come to an end. From left to right and top to bottom, the panels correspond to sectors 35, 37, 39, 41, 43, and 45.}
   \label{fig:Comparison}
   \end{figure}

\subsection{Plane-of-Sky Velocity}
\label{sec:velocity-pos}

We apply the same procedure as in \sect{velocity-disk} to obtain curves representative of the evolution of the shock fronts in the low corona, \ie~ in the plane-of-sky frame. The DT maps from AIA 193\,\AA~sectors 35, 37, 39, 41, 43, and 45 are displayed in \fig{Comparison} with the average splines obtained via the visual method superimposed in white. For comparison, we also determine the splines from H$\alpha$ and AIA 304\,\AA~in the plane-of-sky frame. These are represented by the yellow and red lines respectively in the same figure.

We obtain the kinematic characteristics of the shock wavefront in each sector for AIA 193\,\AA~in the same way as in \sect{velocity-disk}, \ie~by fitting 
power-law curves to the average splines in the full temporal range. 
The results are presented in Table~\ref{table193}, calculating the instantaneous values of acceleration and speed for the times  t$_{on}$=17:45:16 UT and t=17:50:00 UT.

\begin{table}
\caption{Accelerations and speeds derived from 193\,\AA~observations using a power-law fit.}
\label{table193}
\begin{tabular}{ccccc} 
\hline
Sect  & \multicolumn{2}{c}{acceleration [\kmss]}& \multicolumn{2}{c}{velocity [\kms]} \\
  193\,\AA      &  t$_{on}$=17:45:16 UT & t=17:50:00 UT & t$_{on}$=17:45:16 UT & t=17:50:00 UT \\
  \hline
16  &  -0.235 $\pm$ 0.012  &  -0.162 $\pm$ 0.008  &  ~589.5 $\pm$ ~35.1  &  ~534.2 $\pm$ ~32.4 \\
17  &  -0.389 $\pm$ 0.010  &  -0.258 $\pm$ 0.007  &  ~687.5 $\pm$ ~20.3  &  ~597.6 $\pm$ ~17.9 \\
18  &  -0.335 $\pm$ 0.007  &  -0.225 $\pm$ 0.005  &  ~651.2 $\pm$ ~16.5  &  ~573.5 $\pm$ ~14.7 \\
19  &  -0.349 $\pm$ 0.008  &  -0.235 $\pm$ 0.005  &  ~682.2 $\pm$ ~17.6  &  ~601.1 $\pm$ ~15.8 \\
20  &  -0.455 $\pm$ 0.008  &  -0.302 $\pm$ 0.005  &  ~812.1 $\pm$ ~16.0  &  ~707.1 $\pm$ ~14.1 \\
21  &  -0.573 $\pm$ 0.008  &  -0.370 $\pm$ 0.005  &  ~844.5 $\pm$ ~13.3  &  ~713.9 $\pm$ ~11.4 \\
22  &  -0.558 $\pm$ 0.011  &  -0.363 $\pm$ 0.007  &  ~856.5 $\pm$ ~18.2  &  ~728.8 $\pm$ ~15.7 \\
23  &  -0.422 $\pm$ 0.007  &  -0.286 $\pm$ 0.005  &  ~888.2 $\pm$ ~17.3  &  ~789.6 $\pm$ ~15.6 \\
24  &  -0.419 $\pm$ 0.010  &  -0.285 $\pm$ 0.007  &  ~912.9 $\pm$ ~25.9  &  ~814.9 $\pm$ ~23.5 \\
25  &  -0.416 $\pm$ 0.008  &  -0.285 $\pm$ 0.005  &  ~956.2 $\pm$ ~20.5  &  ~858.7 $\pm$ ~18.7 \\
26  &  -0.432 $\pm$ 0.008  &  -0.296 $\pm$ 0.006  &  ~999.1 $\pm$ ~21.7  &  ~897.7 $\pm$ ~19.8 \\
27  &  -0.412 $\pm$ 0.009  &  -0.283 $\pm$ 0.006  &  ~976.5 $\pm$ ~25.1  &  ~879.7 $\pm$ ~22.9 \\
28  &  -0.465 $\pm$ 0.008  &  -0.318 $\pm$ 0.006  &  1045.0 $\pm$ ~21.3  &  ~936.0 $\pm$ ~19.4 \\
29  &  -0.500 $\pm$ 0.010  &  -0.342 $\pm$ 0.007  &  1144.9 $\pm$ ~26.2  &  1027.6 $\pm$ ~23.9 \\
30  &  -0.524 $\pm$ 0.008  &  -0.360 $\pm$ 0.005  &  1217.6 $\pm$ ~20.6  &  1094.5 $\pm$ ~18.8 \\
31  &  -0.534 $\pm$ 0.001  &  -0.366 $\pm$ 0.001  &  1241.4 $\pm$ ~~3.9  &  1116.0 $\pm$ ~~3.6 \\
32  &  -0.527 $\pm$ 0.008  &  -0.361 $\pm$ 0.006  &  1214.6 $\pm$ ~22.8  &  1090.9 $\pm$ ~20.8  \\
33  &  -0.550 $\pm$ 0.007  &  -0.375 $\pm$ 0.005  &  1205.2 $\pm$ ~17.5  &  1076.4 $\pm$ ~15.9 \\
34  &  -0.683 $\pm$ 0.008  &  -0.445 $\pm$ 0.005  &  1070.7 $\pm$ ~13.2  &  ~914.3 $\pm$ ~11.4 \\
35  &  -0.566 $\pm$ 0.007  &  -0.376 $\pm$ 0.005  &  1001.2 $\pm$ ~14.3  &  ~870.5 $\pm$ ~12.6 \\
36  &  -0.571 $\pm$ 0.005  &  -0.382 $\pm$ 0.004  &  1071.1 $\pm$ ~11.6  &  ~938.7 $\pm$ ~10.3 \\
37  &  -0.494 $\pm$ 0.006  &  -0.336 $\pm$ 0.004  &  1077.6 $\pm$ ~14.0  &  ~962.1 $\pm$ ~12.7 \\
38  &  -0.494 $\pm$ 0.005  &  -0.336 $\pm$ 0.004  &  1067.4 $\pm$ ~13.2  &  ~951.8 $\pm$ ~11.9 \\
39  &  -0.577 $\pm$ 0.007  &  -0.387 $\pm$ 0.005  &  1102.1 $\pm$ ~14.6  &  ~968.3 $\pm$ ~13.0 \\
40  &  -0.679 $\pm$ 0.005  &  -0.444 $\pm$ 0.003  &  1083.4 $\pm$ ~~8.8  &  ~927.8 $\pm$ ~~7.6 \\
41  &  -0.512 $\pm$ 0.004  &  -0.343 $\pm$ 0.002  &  ~963.6 $\pm$ ~~7.7  &  ~844.8 $\pm$ ~~6.9 \\
42  &  -0.533 $\pm$ 0.004  &  -0.351 $\pm$ 0.003  &  ~883.8 $\pm$ ~~7.2  &  ~761.1 $\pm$ ~~6.3 \\
43  &  -0.518 $\pm$ 0.006  &  -0.339 $\pm$ 0.004  &  ~830.8 $\pm$ ~10.2  &  ~712.0 $\pm$ ~~8.9 \\
44  &  -0.546 $\pm$ 0.003  &  -0.353 $\pm$ 0.002  &  ~808.6 $\pm$ ~~5.6  &  ~684.1 $\pm$ ~~4.8 \\
45  &  -0.370 $\pm$ 0.005  &  -0.246 $\pm$ 0.003  &  ~675.1 $\pm$ ~10.0  &  ~589.6 $\pm$ ~~8.9 \\
46  &  -0.446 $\pm$ 0.007  &  -0.295 $\pm$ 0.005  &  ~770.9 $\pm$ ~14.0  &  ~668.1 $\pm$ ~12.3 \\
47  &  -0.381 $\pm$ 0.031  &  -0.267 $\pm$ 0.022  &  1109.0 $\pm$ 110.6  &  1018.5 $\pm$ 103.2 \\
mean&  -0.483 $\pm$ 0.097  &  -0.324 $\pm$ 0.062  &  ~951.3 $\pm$ 181.4  &  ~839.1 $\pm$ 168.3 \\
\hline
\end{tabular}
\end{table}

  \begin{figure}    
   \centerline{
	             \includegraphics[width=1.0\textwidth]{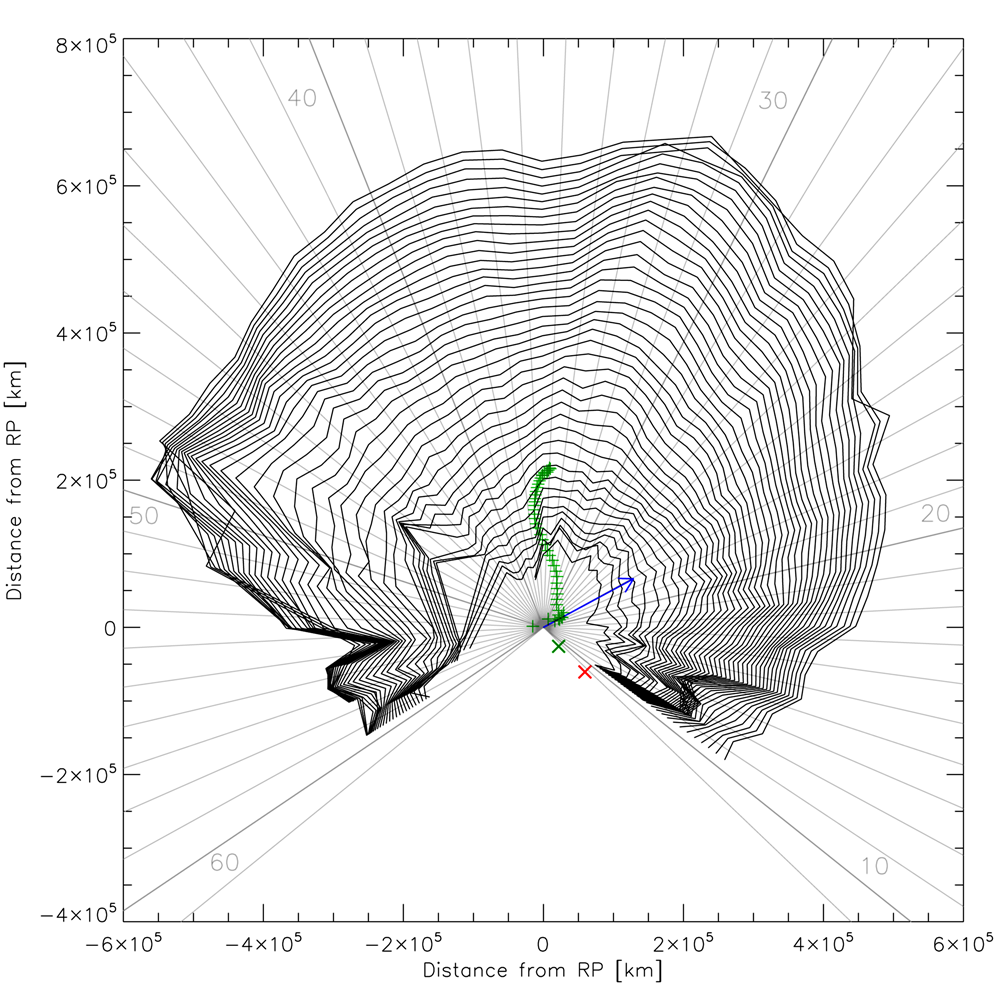}
              }
              \caption{An $(x,y)$ plot showing the angular dependence of the AIA 193\,\AA~wave evolution in the plane-of-sky view. The wavefronts are obtained from the set of points determined from the average splines for each sector. Some sectors have been labeled. Sectors depart from the location originally defined as the RP. The curves are traced every 15 s, from t$_{on}$=17:45:16 UT to t=17:55:46 UT. The green `+' signs indicate the locations of the centers of every wavefront. The blue arrow is the projection on the plane of view of a vector normal to the surface at the RP coordinates. The `x' signs indicate the extrapolated centers of the circumferences to the reference flare time, with green corresponding to a linear fit and red to a quadratic fit. }
   \label{fig:xyWaveFronts}
   \end{figure}

\subsection{Coronal Wave Model}
\label{sec:model}

To investigate the angular dependence of the wave evolution in the plane-of-sky view, we build an $(x,y)$ plot based on the shock-front curves obtained from AIA 193\,\AA~data. In this plot, increasing $x$ corresponds to the east--west direction and increasing $y$ to the south--north direction. This graph is obtained by plotting the $d_s$ distances   every 15 s from the RP, derived from the average splines along the corresponding trajectories previously used to define the sectors,  from t$_{on}$=17:45:16 UT to t=17:55:46 UT. The resulting wavefronts can be seen in \fig{xyWaveFronts}. Several characteristics of the shock evolution are evident:

\begin{itemize}
\item{A fairly homogeneous circular pattern is observed between sectors 16 and 45, which would indicate the evolution of a quasi-circular coronal front in the plane-of-sky view. This front could be interpreted as the borderline of a quasi-spherical 3D-dome EUV wave evolving towards the observer. } 
\item{Between sectors 29 and 34, however, the wavefront exhibits a lobe, suggesting that it moves faster in this region than in the remaining sectors. This is in agreement with the initial speed values listed in Table~\ref{table193}.}
\item{Southward from AR 12017 no shock wave is detected, probably because the high Alfv\'en velocity in the AR inhibits the shock formation and propagation \citep{Vrsnak08}. To both sides of the AR (sectors 10\,--\,15 and 49\,--\,59) the shock exhibits an irregular pattern. This is probably the result of a more diffuse and less intense shock front in these sectors, thus implying higher measurement errors.}
\item{The entire circular pattern appears to move northwards as indicated by the green plus signs that denote the locations of the centers of every wavefront. These are determined by interpolating circumferences to the wavefronts (see the following paragraphs), taking into account all sectors between 16 and 45.}
\item{The bulk of the 3D-dome EUV wave does not follow the radial direction, as evidenced by the blue arrow, that indicates the plane-of-view projection of a radial vector located at the RP.}

\end{itemize}

To obtain the general characteristics of the shock evolution, we perform the  interpolation of circumferences to the shock fronts that evolve in time as displayed in \fig{xyWaveFronts}. For the fitting we use a Levenberg-Marquardt robust algorithm from the MPFIT libraries \citep{Markwardt09}.

\begin{figure}    
   \centerline{
	            \includegraphics[width=1.0\textwidth,clip=]{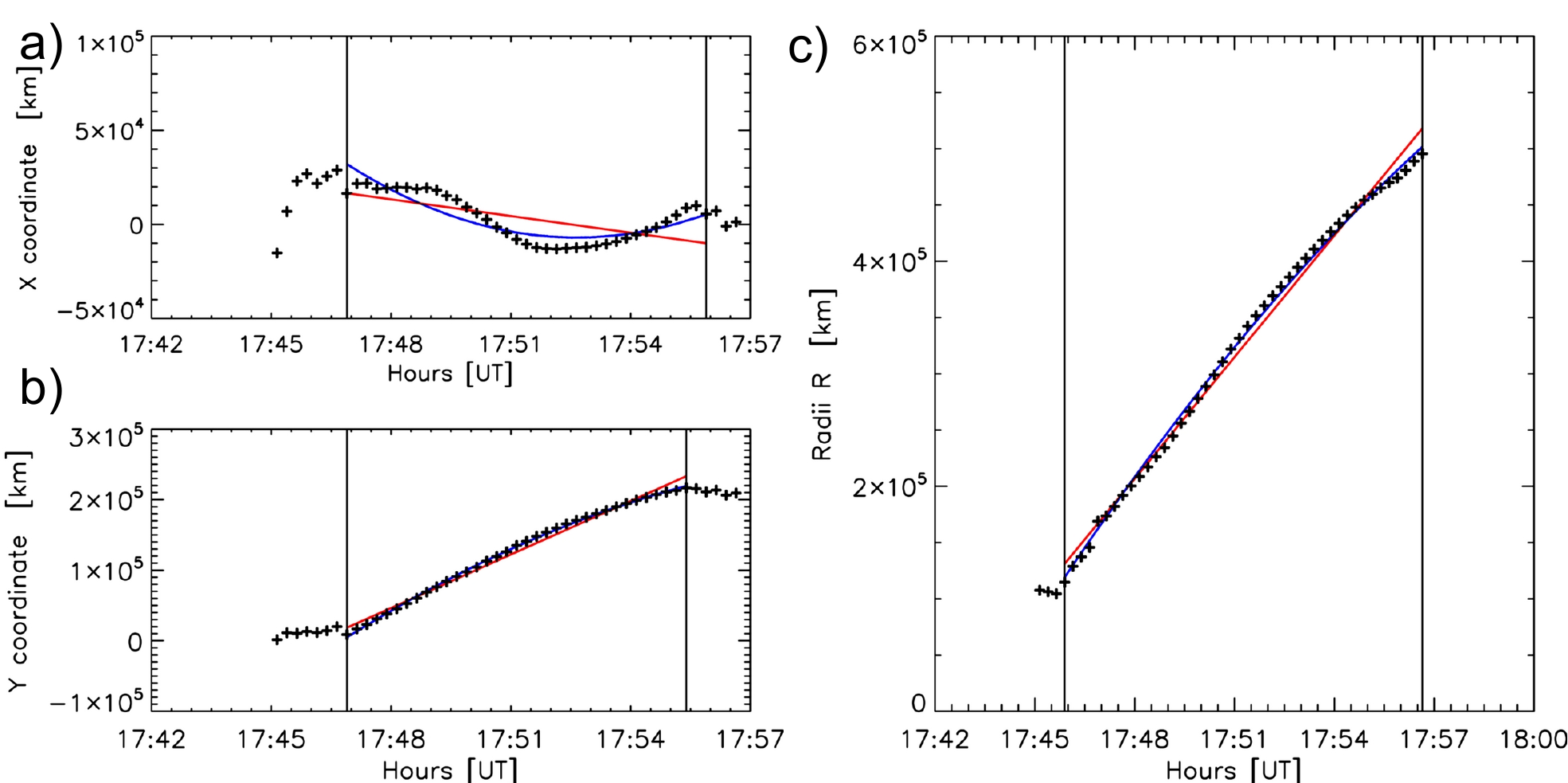}
							}
              \caption{Temporal evolution of the parameters deduced from interpolating circumferences to the wavefronts including sectors between 16 and 45. (a) Corresponds to the $x$-coordinate of the centers of the circumferences. (b) Shows the $y$-coordinate of the centers. (c) Corresponds to the radii of the circumferences. In all panels the vertical black lines delimit the region where the points have been fitted by curves. The red and blue lines represent the linear and quadratic fits, respectively.}
   \label{fig:x-y-radius}
   \end{figure}

The results of the interpolation can be observed in \fig{x-y-radius}. Panel a shows the evolution in time of the $x$ coordinate of the inferred circumferences' centers. The red and blue lines in this panel are the linear and quadratic fits to the points within the range indicated by the vertical lines neglecting the discordant data points at the beginning and end times. The evolution of the $y$ coordinate of the circumferences' centers is similarly shown in panel b, where a fast displacement in this direction is evident. Panel c shows the evolution of the circumferences' radii, which increase with a speed of $v_r \approx 600$\,\kms. The results of the fitting are listed in Table~\ref{tablecircumferences}. 

The centers of the interpolated circumferences are superimposed in \fig{xyWaveFronts} as green plus signs.
They show a pronounced northward drift as time evolves, slightly oscillating in the east--west direction. 

To obtain an indication of the origin site of the wave event, we perform a backwards extrapolation to the flare reference time [t$_{on}$] of the $x$ and $y$ values, using the linear and quadratic fits of \fig{x-y-radius}a and b. The results can be observed in \fig{xyWaveFronts}, where the green and red `x' signs indicate the $(x,y)$ position obtained from the linear and quadratic fits, respectively.  These two points do not coincide with the RP, but lie in the close vicinity of the AR.

\begin{table}
\caption{Coefficients for the linear (b and c) and quadratic (a, b, and c) fitting of the circumferences' centers and radii with time.}
\label{tablecircumferences}
\begin{tabular}{lccc} 
  \hline
fit	& a [\kmss] & b [\kms] & c [km] \ \ \\
  \hline
$x$ linear          &  -----     & -4.923 $\times$ 10$^{1}$   &  ~3.168 $\times$ 10$^{6}$  \\
$x$ quadratic       &  ~0.327    & -4.213 $\times$ 10$^{4}$   &  ~1.356 $\times$ 10$^{9}$  \\
$y$ linear          &  -----     & ~4.211 $\times$ 10$^{2}$   &  -2.693 $\times$ 10$^{7}$  \\
$y$ quadratic       &  -0.328    & ~4.258 $\times$ 10$^{4}$   &  -1.382 $\times$ 10$^{9}$  \\
radius linear       &  -----     & ~6.001 $\times$ 10$^{2}$   &  -3.825 $\times$ 10$^{7}$  \\
radius quadratic    &  -0.219    & ~2.870 $\times$ 10$^{4}$   &  -9.410 $\times$ 10$^{8}$  \\

\hline
\end{tabular}
\end{table}

Next, we correlate the measured coronal shock with the chromospheric Moreton traces. Several authors suggest that Moreton waves can be explained by a 3D dome-shaped coronal shock front pushing down and sweeping the chromosphere 
(see references in \sect{intro}). If this is the case, we should expect that both measurements, chromospheric and coronal, coincide when intersecting the coronal shock-front with the Sun's surface.

 Following this hypothesis the results obtained above  by measuring the shock fronts (see \figs{xyWaveFronts}{x-y-radius})  
could be assumed as the borderline in the plane-of-sky of a 3D structure, more or less regular, that propagates outward in the corona. Furthermore,  this structure probably drifts northward, according to the 
temporal evolution of the $(x,y)$ coordinates of the interpolated circumferences' centers. 

We assume a simple model for this 3D structure, namely an expanding sphere whose center drifts with constant velocity [${\vec v}=(v_x, v_y, v_z)$] and starts moving from
a point located at a certain height [$h$] above the RP. This model implicitly assumes a piston driver for the wave event. This piston driver is initially located in the vicinity of the AR and starts moving along a specific 3D path at a particular time [$t_0$], close to the flare onset time [$t_{on}$], compatible with the 
filament ejection and CME lift off (see \eg~\opencite{Liu15} for a detailed analysis of the first stages of the event and \sect{intro} for a summary). 

  \begin{figure}    
   \centerline{
	             \includegraphics[width=1.\textwidth]{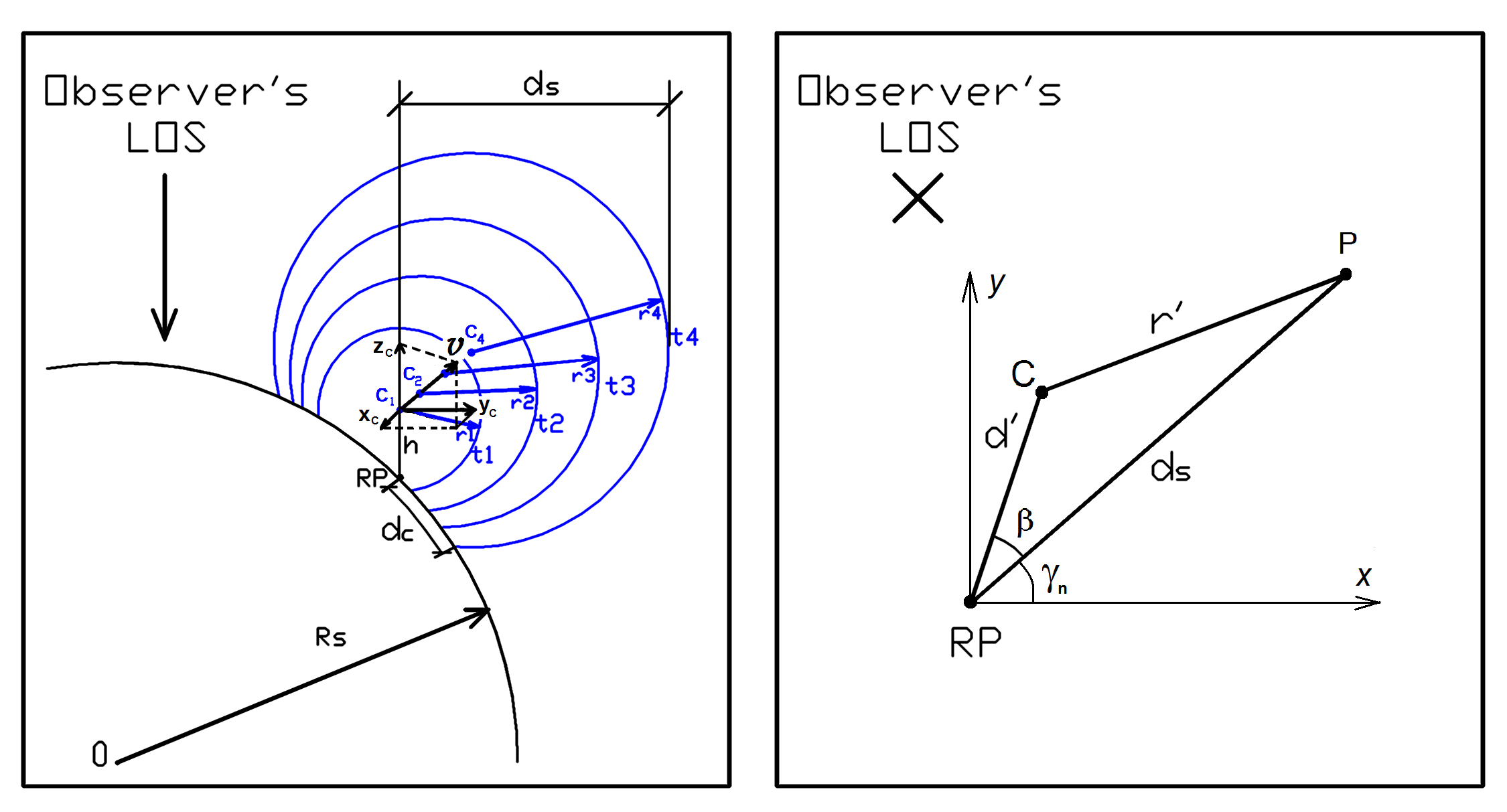}
              }
              \caption{a) Sketch of the coronal wave-model showing the relationship between an expanding and outward-moving sphere and its chromospheric trace as it evolves in time. $Rs$ stands for the solar radius and RP for the radiant point, $d_c$ is the distance from the RP to the chromospheric trace, $d_s$ is the distance from the RP to the projection of the sphere on the plane-of-view in the direction of a particular sector, and $h$ is the height of the sphere's center. (b) Scheme of the plane-of-view used to calculate the measured radius of the expanding sphere [$r'$] (see \eq{cose}).}
   \label{fig:Scheme}
   \end{figure}

A sketch of this model can be seen in \fig{Scheme}a. This panel shows 
a planar cut of the solar sphere along a great circle that includes the RP. The observer's LOS is indicated in the figure. The observer is able to measure both the plane-of-sky coronal distances between the shock and the RP [$d_s$] and the chromospheric distances traveled by the Moreton traces over the solar surface [$d_c$]. 

If the Moreton trace is originated by the shock wave sweeping the chromosphere, the distance $d_c(t_i)$ must match the intersection of a sphere centered at the point $C(t_i)=\left(x_c(t_i),y_c(t_i), z_c(t_i)\right)$ and with radius $r(t_i)$. 
We use $t_i$ instead of $t$ to make reference to the specific instants of time we use to define the average splines in the plane-of-sky and surface frame. 

Due to inhomogeneities in the shock front in the plane-of sky frame seen in \fig{xyWaveFronts},  we use a measured sphere radius 
[$r'(t_i)$] obtained from the distances [$d_s(t_i)$],
instead of the mean radius [$r(t_i)$] obtained from the geometrical interpolation of \fig{x-y-radius}c.  
Furthermore, $d_s(t_i)$ corresponding to sector $n$ in the plane-of-sky frame, must agree with $d_c(t_i)$ of sector $n$ in the surface frame, at the same time [$t_i$]. 
Each radius [$r'$] can be obtained from the corresponding distance [$d_s$] as it can be deduced from the scheme in \fig{Scheme}b, which exhibits an $(x,y)$ plane-of-sky view of the measured distance [$d_s$] of a point $P$, the RP, and the center of the expanding spheres, all considered at the same time $t_i$. This view would correspond to one of the images captured during the event with the observer's located in front of the figure.
According to \fig{Scheme}b,  if the origin is set in coincidence with the RP coordinates, from the law of cosines we obtain:

   \begin{equation}  \label{eq:cose}
     r' = \sqrt{d'^2 + d_s^2 - 2 d' d_s \cos \beta },
   \end{equation}

\noindent	
where
	
	   \begin{equation}  \label{Eq-dprim}
     d' = \sqrt{C_x^2 +  C_y^2 }
   \end{equation}
	
\noindent
and the angle $\beta$ can be obtained as:

	   \begin{equation}  \label{Eq-beta}
     \beta = \arctan \frac{ C_y}{ C_x} - \gamma_n,
   \end{equation}

\noindent	
where $\gamma_n$ is the angle with respect to the $x$ axis of the analyzed sector $n$ and ($C_x$,$C_y$) are the coordinates of the center $C$.
To compute the intersection of the measured shock sphere with the solar surface, we design an algorithm that finds which of the points $P_j(t_i)=(x_j(t_i), y_j(t_i), z_j(t_i))$ of the set of distances [$d_c(t_i)$]  and sector $n$ traced over the solar sphere, \ie~in the surface frame, fits the expanding sphere of radius $r'(t_i)$, obtained as explained above and centered at point $C(t_i)=(x_c(t_i),y_c(t_i),z_c(t_i))$. 
The coordinates $x_c(t_i)$, $y_c(t_i)$, $z_c(t_i)$ of the center point $C(t_i)$ are obtained as:

\begin{eqnarray} 
x_c(t_i) &=&  v_x  \Delta_t + RP_x,   \\
y_c(t_i) &=&  v_y  \Delta_t + RP_y,  \\
z_c(t_i) &=&  v_z  \Delta_t + RP_z + h, 
\label{Eq-center}
\end{eqnarray}

\noindent
where $\Delta_t = (t_i-t_0)$.

The algorithm finds the points $P_j(t_i)$ in sector $n$ from the following inequation:

	   \begin{equation}  \label{Eq-minimum}
        [ x_j(t_i) - x_c(t_i) ]^2 +  [ y_j(t_i) - y_c(t_i) ]^2 + [ z_j(t_i) - z_c(t_i) ]^2 - r(t_i)^2 \ \ {\rm <} \ \ \epsilon       
   \end{equation}
	
\noindent
for $i=1,\ldots l$ and $j=1,\ldots m$, where $\epsilon$ is the second power of an infinitesimal distance, $l$ is the time of the measurement in seconds and $m$ the number of points for path $n$, in this case 1000.

The free parameters of the model are the values $h$, $v_x$, $v_y$, $v_z$, and the time $t_0$. We estimate the velocities $v_x$ and $v_y$ from the linear fit parameters listed in Table~\ref{tablecircumferences}. We obtain $t_0$ as the average value of the roots of the interpolated quadratic curves of the average splines of sectors 16 to 47 and find that $t_0=$\,17:44:06.75 UT.

\cite{Kleint15} measured the velocity of the ejected filament associated with the 29 March flare. They estimated a range of projected speeds from 130 to 230 \kms~at 17:44:37 UT and 340 to 700 \kms~at 17:45:13 UT. The initial filament height could not be measured by these authors due to the low Doppler signal, implying that such a low speed is characteristic of chromosphere and transition region heights. In consequence, we assume $h$ = 1000 km.
As mentioned in \sect{intro}, \cite{Liu15} found an asymmetric filament eruption and determined a speed of 620 \kms~ at around 17:43 UT.

For the remaining free parameter [$v_z$] we find an acceptable fit choosing 310 \kms~when varying $v_z$ in steps of 10 \kms.

 \begin{figure}    
   \centerline{
	             \includegraphics[width=1.0\textwidth]{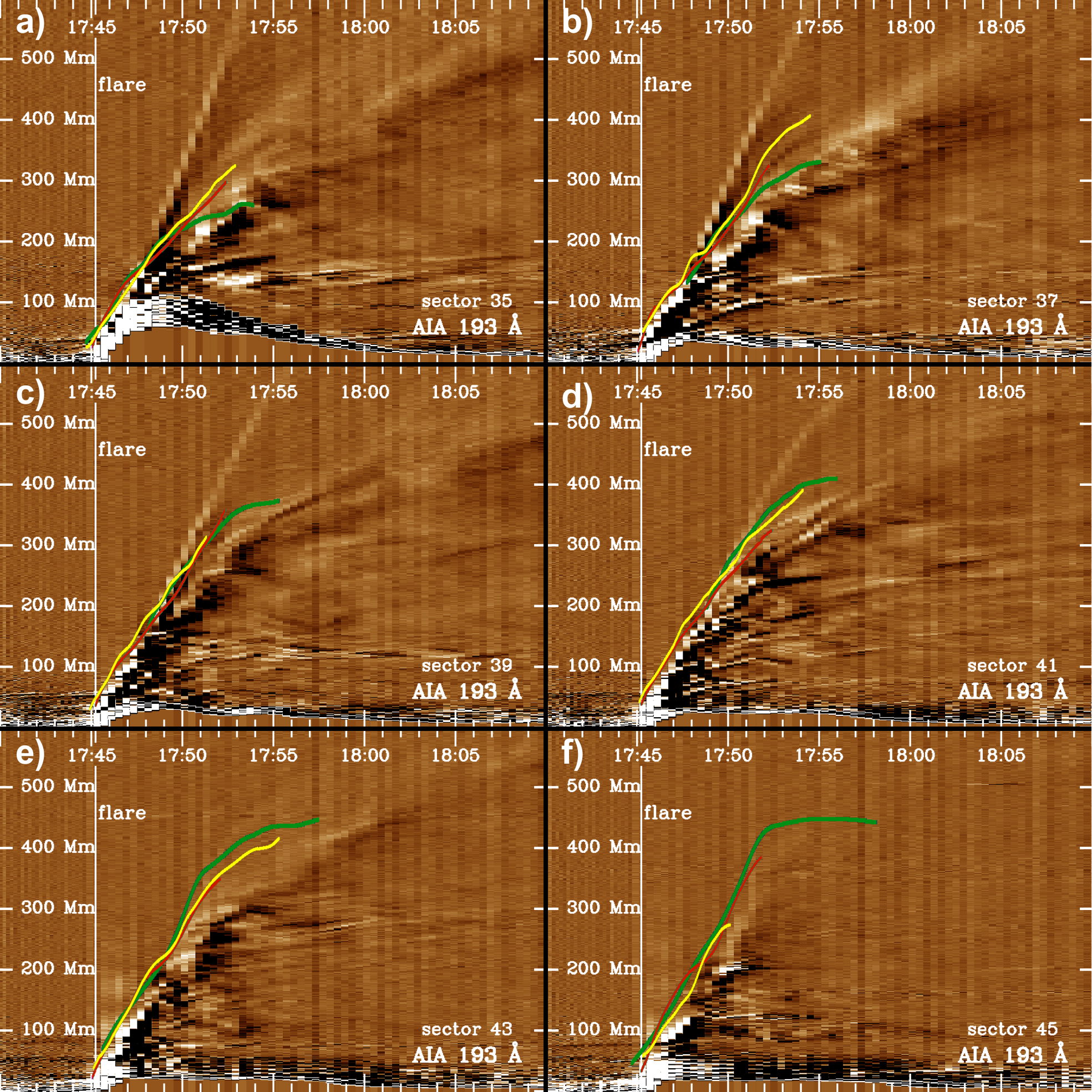}
              }
              \caption{DT maps determined from AIA 193\,\AA~data in the surface frame with the results from applying the coronal-wave model to a set of sectors superimposed. The green line is the modeled chromospheric trace of the fastest shock-wavefront seen 
in the AIA 193\,\AA~line, while the yellow and red lines are the profiles of the wavefront in the H$\alpha$ and He\,{\sc ii} lines, respectively. Panels a to f correspond to sectors 35, 37, 39, 41, 43, and 45 (from left to right and top to bottom).}
   \label{fig:ModelRes}
   \end{figure}

Finally, we find a reasonable fit of our model when the sphere's center is moving at a constant speed with components $v_x=-49.23$\,\kms, $v_y=421.05\,$\kms, and  $v_z=310.0$\,\kms.  

In \fig{ModelRes}, we take the DT maps built from AIA 193\,\AA~for several sectors in the surface frame as a basis to draw various profiles. We superimpose to these DT maps the H$\alpha$ and AIA 304\,\AA~measured splines in yellow and red, respectively. These roughly delineate the behavior of the brighter front that travels behind the faint and faster shock-wave seen in the coronal lines. The results of the model, \ie~the deduced shock-wave intersection at the solar surface, is drawn in green. From the figure, we see a general good fit between our model and the H$\alpha$ and He\,{\sc ii} traces, which suggests that the intersection of these spheres with the chromosphere corresponds in fact with the Moreton and He\,{\sc ii} waves. In the western sectors 35 and 37 and beyond $\approx$\,200\,Mm of the RP, the model results appear below the chromospheric traces, \ie~with slower surface speed (see Figures\ref{fig:ModelRes}a\,--\,b), while the opposite applies for sector 43. This discrepancy is addressed in \sect{regimes}.

 \begin{figure}    
   \centerline{
	             \includegraphics[width=1.\textwidth]{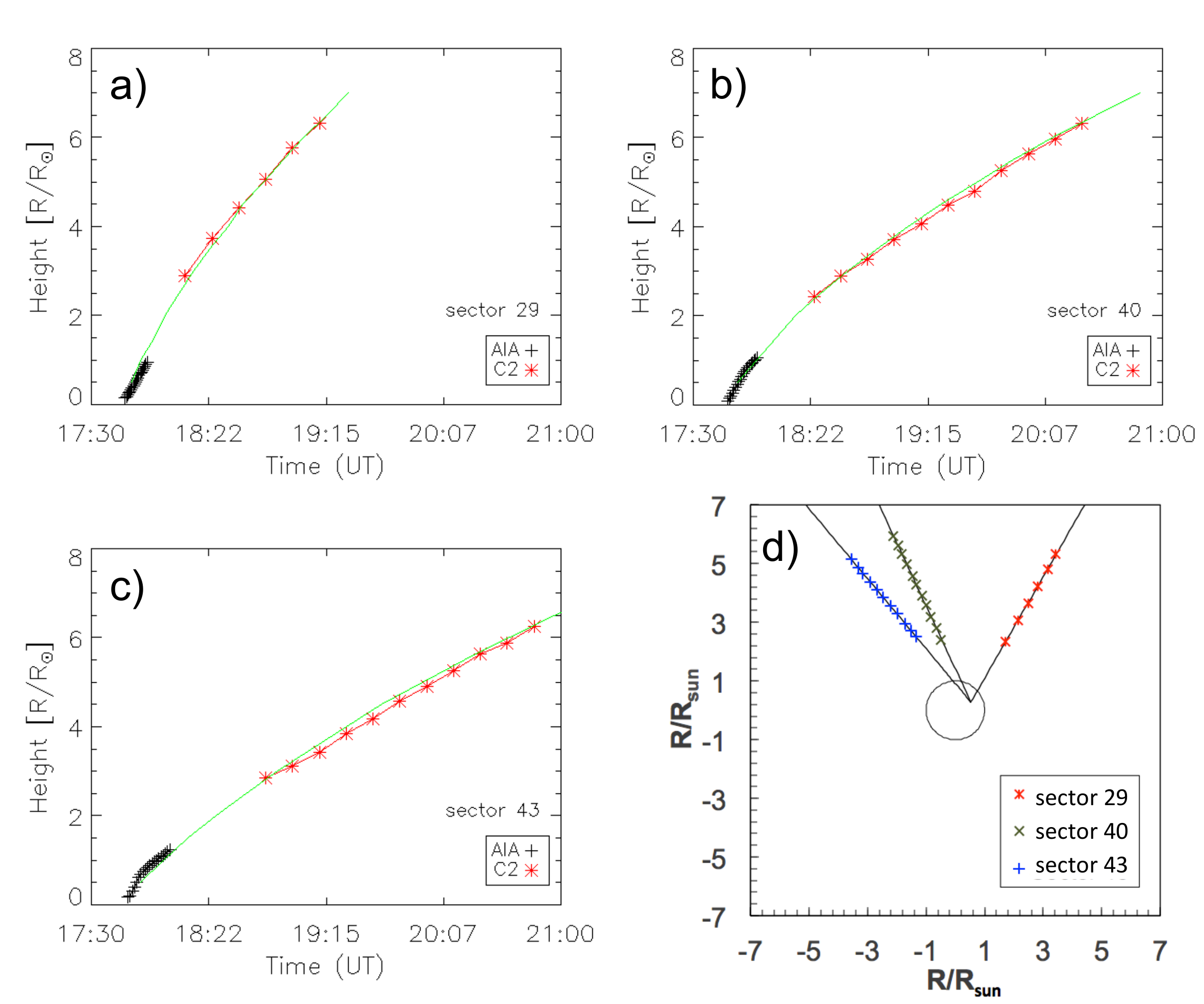}
              }
              \caption{a) Distance-time plot corresponding to sector 29 with black plus signs representing plane-of-sky points derived from AIA 193\,\AA~and red asterisks plane-of-sky measurements of the CME's shock. (b) Same as panel a but for sector 40. (c) Same as panel a but for sector 43. (d) An ($x,y$) plot indicating the three directions along which the distance-time measurements of the CME shock have been performed in LASCO C2 images, as well as the measured points. The circle represents 1\,\Rsun.}
   \label{fig:CME2}
   \end{figure}

\subsection{Correspondence between EUV and White-light Features}
\label{sec:link-WL}

To understand the origin of the 3D-dome EUV wave
 observed in the low corona in connection with the white-light counterpart detected in coronagraphic images, we build DT plots along three different directions. Both the pathway of the shock wave in the low corona and that of the CME's shock are shown together in \fig{CME2} along the central direction of sectors 29 (panel a), 40 (panel b), and 43 (panel c). In the low corona, the shock front is tracked using the DT maps constructed from AIA 193\,\AA~and under the plane-of-sky assumption, as explained in \sect{velocity-pos} (see also \figs{Stack193}{Comparison}) and considering the RP as the origin of the wavefront. We track the shock in LASCO C2 images along the same central directions of sectors 29, 40, and 43, which are shown as solid lines in \fig{CME2}d. In this panel the symbols on the lines mark the positions of the shock fronts measured at different times for the three sectors. The measurements are performed under the plane-of-sky assumption as well. 

The DT points derived from the low and white-light coronal data are fitted using $d(t)=\sqrt{(at+b)}+c$, where the fit to the data is performed applying the same Levenberg-Marquardt least-square approximation as in \sect{model}. This equation was also applied by \cite{Cremades15} to fit points of a distance-time plot of a CME/shock that resulted from the combination of white-light corona, interplanetary type II radio, and {\it in situ} data. The fit is very good both in correspondence with the fast growth of the height in the low corona and also with the gradual increase registered later in the white-light corona. A quadratic equation was also applied to fit the data, but the fit was poor.

\section{Discussion and Conclusions}
\label{sec:discussion}

We analyze the Moreton-wave event on 29 March 2014 observed in H$\alpha$ by HASTA using a 5-second temporal cadence, together with high-resolution transition region and coronal EUV observations from AIA, which enables the analysis of wave phenomena in different regimes of the solar atmosphere. We focus our study on the spatial behavior of the waves, through a detailed analysis of the angular evolution of the wavefronts, to find correspondences between the features observed in different bands and atmospheric layers. 
In this section we address the kinematics of the event, the correspondence found between the features detected in the various regimes, and the origin of the shocks within the blast-wave and piston-driven scenarios.

\subsection{About the Kinematics of the Wave Event}
\label{sec:kinematics}

We firstly obtain the kinematic parameters of H$\alpha$ and He\,{\sc ii} band wavefronts, both measured on the curvature of the solar surface (\sect{velocity-disk}). The parameters are calculated considering an RP located at the 
brightest flare kernel, \ie~not using an RP derived from tracing arcs of circumferences over the chromospheric H$\alpha$ fronts as done by other authors \citep{Warmuth04a,Warmuth04b,Francile13}. Despite the differences between the methods of measurement, we consider that the values of the kinematic parameters (distances, speeds, and accelerations) should be similar far from the AR. 

The first H$\alpha$ traces are detected $\approx$\,50\,Mm away from the brightest flare-kernel site. This distance to the flare is shorter than the mean distance determined by \citet{Warmuth04b} for several Moreton events. The angular extent of the chromospheric disturbance is $\approx$\,80\,$\degree$. The H$\alpha$ kinematic parameters show the typical characteristics of Moreton chromospheric waves, with initial speeds between 650\,--\,900\,\kms~decaying to 490\,--\,800\,\kms~five minutes later. The instantaneous accelerations determined at the flare onset time and five minutes later listed in Table~\ref{tableHa} yield a mean initial deceleration of $\approx$\,0.365 \kmss~that decreases in absolute value with time. 
The analysis by sectors reveals dispersions in the speed and acceleration values, suggesting a non-homogeneous angular behavior of the Moreton wavefronts. A similar kinematic analysis performed on the He\,{\sc ii} distance-time maps shows similar values of acceleration and speed with respect to H$\alpha$. This suggests that the upper chromosphere and the transition region respond similarly to the increase of pressure or density produced by the arrival of the coronal shock which leads to emission enhancements \citep{Vrsnak08, Leenaarts12, Krause15}.

The kinematic parameters of the shock 
front in the AIA 193\,\AA~band are subsequently determined for all sectors. A directional dependence of the acceleration and speed is evident, with the highest deceleration corresponding to the zone covering sectors  28\,--\,44, \ie~the northward direction along which the Moreton fronts are observed. The visibility range could contribute to this effect given that the wave can be tracked to much larger distances for these sectors, thus allowing for the deceleration to become significant in the power-law fit. 
The speeds calculated $\approx$\,5\,min after the wave ignition are in general greater than 900 \kms~in sectors 28\,--\,40. 
Typical coronal values in quiet regions are $c_s = 185$ \kms~for the sound speed, $v_A = 273$ \kms~for the Alfv\'en speed, and $v_{ms}=330$ \kms~for the fast magneto-acoustic speed \citep{Warmuth15}, implying for our wave a magneto-acoustic Mach number of $\approx$ 3, considering the shock evolving close to the solar surface, toward the north, and far from the AR. High mean deceleration rates correspond to high Mach-number values of a shock wave, whereby the wave losses energy rapidly and decelerates faster. 
Regardless of this fact, the speed of the wave northwards and far from the AR is still significant enough to ensure a high compression ratio at chromospheric levels to generate the Moreton wave. On the other hand, a lower Mach number is expected in the vicinity and above the AR. 

\begin{figure}    
   \centerline{
	             \includegraphics[width=0.95\textwidth]{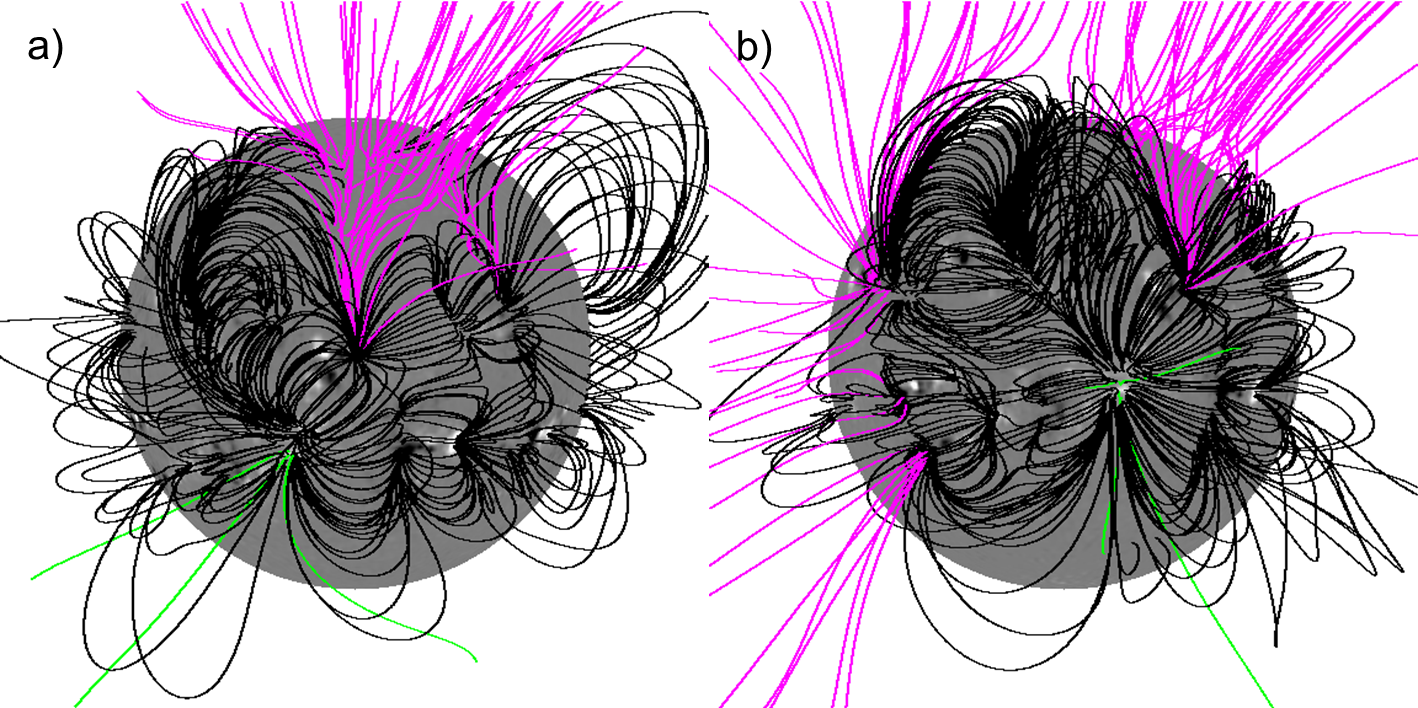}
              }
              \caption{PFSS model for Carrington rotation 2148. (a) Shows the AR at central meridian passage. (b) Corresponds to the AR as viewed from Earth's perspective on 29 March 2014 at 12:04:00 UT. The field-line color
convention is such that black indicates closed lines and magenta (green) corresponds to open lines anchored in the negative-polarity (positive-polarity) field.
} 

   \label{fig:PFSS}
   \end{figure}

The wave brightness is noticeably higher between sectors 35\,--\,43 in H$\alpha$ and He\,{\sc ii} (see \fig{StackSurf}).  
The fastest wavefront detected in the AIA 193\,\AA~band comprises sectors 28\,--\,40.
This partial overlap suggests a relationship between the speed of the coronal shock and the  preferential propagation direction of the Moreton wave. The higher speed of the coronal shock in this direction 
can be attributed mainly to a special characteristic of the coronal medium and the magnetic field configuration through which the disturbance propagates, but also the energy supplied initially by the piston driver could account for this effect \citep{Krause15}. \fig{PFSS} shows the Potential Field Source Surface (PFSS) model for Carrington rotation 2148 with the
AR at central meridian passage (\fig{PFSS}a) and as viewed from Earth's perspective on 29 March 2014 (\fig{PFSS}b). A region of open magnetic field lines can be seen in correspondence with sectors 35\,--\,43, which sets favorable conditions for the propagation of a coronal MHD shock
able to generate Moreton disturbances \citep{Zhang11}. Furthermore, the shock speed is fundamentally determined by the coronal magnetic field value, but the compression ratio decreases with increasing magnetic field \citep{Krause15}. Thus, the ``field-line valley'' to the north of AR 12017 having low Alfv\'en speed implies a region of high compression ratio, consequently a shock wave traveling through it can produce a strong compression over the chromosphere and eventually a Moreton signature. This is in agreement with the results found by \citet{Zhang11}, who analyzed the magnetic field configuration present in 13 Moreton events and concluded that the waves mostly propagate either in regions of large-scale closed magnetic loops or along valleys delimited by two sets of separated magnetic loops. Furthermore, given that
close to the solar surface we expect a lower fast magneto-acoustic speed, the wavefronts will tend to curve downward, favoring compression over the chromosphere \citep{Zhang11}. All these facts support the hypothesis that the chromospheric Moreton fronts in our analyzed event are produced by a coronal shock wave.

A lobe comprising sectors 29\,--\,34 in which the wave speeds are high is seen in \fig{xyWaveFronts}. Following our previous discussion, these higher speed values cannot be attributed to a fast magneto-acoustic speed in the region, since it should be low. This speed asymmetry could also be caused by the shock refracting in the high magnetosonic walls of the valley or by a particular behavior of a hypothetical piston producing the wave in this area.

The non-radial propagation of the CME could be explained by the magnetic configuration to the north of AR 12017. The high magnetic pressure of the AR would force the CME ejection toward the north, where open magnetic field lines and low Alfv\'en speed predominate, without impeding the expansion and propagation of the magnetic structure sustaining the CME.

\subsection{About the Correspondence between the Analyzed Regimes}
\label{sec:regimes}

The stack plots built from the angular sectors for H$\alpha$ and He\,{\sc ii} 
(\fig{StackSurf}) show a strong correlation both in time and surface distances, suggesting that they may respond to the same physical process. Nevertheless, the stack plots appear different when examined in detail, \ie~the H$\alpha$ profiles exhibit brighter traces and  
the He\,{\sc ii} ones show a discontinuous and granular appearance. It should be noted that, in the case of H$\alpha$, the seeing and other common ground-based perturbations strongly affect the data.
In H$\alpha$, regions exhibiting double traces are noticeable (see features indicated by light-blue arrows in sectors 36\,--\,39 of \fig{StackSurf}a), suggesting the presence of more than a single coronal perturbation or a complex physical process that originates the differing chromospheric emissions. Other authors have suggested the presence of more than one Moreton-wavefront in the vicinity of an AR \citep{Muhr10, Francile13}. However, in our case the spatial separation between these two fronts is not large enough to conclude about their presence on a firm basis. This double-trace effect is not visible in the He\,{\sc ii} stack plots.

\fig{Comparison} shows AIA 193\,\AA~DT maps with the superimposed splines traced from AIA 193\,\AA, H$\alpha$ and He\,{\sc ii}, all of them in the plane-of-sky frame, for six representative sectors. 
 The black arrows in \fig{Comparison} denote bright coronal features that appear to persist after the traces derived from H$\alpha$ and He\,{\sc ii} data come to an end.
These features are located at coronal heights, but probably close to the solar surface. They could be caused by the shock front interacting with the lower, denser layers of the corona, when the shock does not have enough speed or intensity to perturb the chromosphere. This suggests that disturbances like near-surface EUV waves have the same physical origin as Moreton waves, \ie~compression in the low corona due to MHD waves or shocks. The low occurrence of Moreton events in relation to near-surface EUV waves can be attributed to the insufficient energy of shock waves to disturb the much denser chromosphere or to the particular characteristics of the coronal environment where they propagate.

The drifting of the center of the circumferences representing the shock-wave evolution (see \figs{xyWaveFronts}{x-y-radius}) inspires a model based on the intersection of a hypothetical spherical shock-front with the solar surface. Dome-shaped coronal waves with an almost circular appearance in EUV and soft X-ray observations have been reported by several authors \citep[\eg][]{Narukage04, Veronig10,  Kozarev11, Ma11,Grechnev11a, Li12, Cheng12}, while numerical simulations also confirm this aspect 
\citep[][and references therein]{Warmuth15}. If the model is accurate enough, the intersection of a spherical shock-front with the solar sphere would coincide with the measured traces in the chromospheric and transition region, sector by sector, if 
a common shock wavefront is responsible for all the effects.   
Combining the results derived from our measurements of the coronal-wave kinematics with the results found by other authors who analyzed the same event (see \sect{model}), the model is able to fit the coronal fronts to the chromospheric ones. The main difficulties of the model arise in the north--west and north--east directions. To the north--west of AR 12017 the model should fit the evolution of shock fronts moving away from the solar surface beyond the solar limb in the low corona, where the fast magneto-acoustic speed of the plasma constantly increases \citep{mann99}; however, as seen in \fig{ModelRes}a and b the model (green line) stays below the chromospheric H$\alpha$ and He {\sc ii} tracings (yellow and red lines, respectively). As mentioned in \sect{kinematics}, the coronal shock-fronts bend toward the solar surface and have a dome shape \citep{Warmuth15}, which implies that the measured distances [$d_s$] will not fit the model's spheres, and therefore the distance [$d_s$] to the west of the RP results in a shorter distance of intersection with the chromosphere (see \fig{ModelRes}a and b). The opposite is true to the east of the RP (see \fig{ModelRes}e and f). This effect is also evident in \fig{xyWaveFronts}, where the westward shock fronts are closer together than those eastward. Some articles report that a lag between the coronal shock-front and the H$\alpha$ chromospheric perturbation of $\approx$\,30 Mm is common \citep{White14, Krause15}. The misalignments in \fig{ModelRes} could be also partially attributed to this effect.

\fig{LowCorona}b and c exhibit running-difference images in the AIA 193\,\AA~and 211\,\AA~bands, where an arc-like coronal feature denoted with green arrows can be seen. This feature propagates behind or jointly with the first shock front, depending on the sector. It is not possible to determine its full appearance or follow its propagation over a considerable period of time; however, we especulate that it may be related to the apparent dual traces visible in sectors 36\,--\,39 of the chromospheric DT-maps (\fig{StackSurf}b). 

In Section~\ref{sec:link-WL} we compare the slope and timings of the fastest AIA 193\,\AA~wavefront with those of the shock ahead of the associated CME propagating in the LASCO C2 field-of-view.
The good agreement evident from \fig{CME2} strongly suggests that the fastest EUV front corresponds to the white-light shock driven by the CME. The misalignment of the two sets of measurements from sector 29 (\fig{CME2}a) may be due to the wave propagating with a large component away from the plane-of-view in this direction, while we are using the plane-of-sky approximation to compare both data sets. The overall correspondence is in agreement with the unified picture of EUV waves synthesized from a series of findings by \citet{Patsourakos12}, whereby the inner brighter front is attributed to the expanding CME loops or bubble and the outer fainter front is the fast-mode wave ahead of the CME flanks and leading edge. Further agreement is found with the findings by \citet{Kwon13}, who noted that the wavefront in the upper (white-light) solar corona turns out to be the counterpart of the EUV wave.

\subsection{ About the Origin of the Shocks} 
\label{sect:origin}

We now discuss how the results of our modeling and of the wave-event analysis fit within 
the blast-wave and piston-driven shock scenarios, \ie~a flare-driven or CME-driven origin.
As discussed above, the directional dependence of the speed can be explained by a spherical shock wavefront whose center moves mostly northward and outward from a RP. This suggests a piston-driven shock likely produced by a CME moving in the direction of the model sphere's center, in which the piston is directly behind the shock front and close to the upper sphere's border.
 In this case, if visible in EUV, the flanks of the piston should appear in the plane of vision as a circular-shaped feature within the ideal spherical shock. In this scenario, the apparent larger speed lobe of sectors 29 - 34 could be an indication of the driver, distorted from the ideal shock sphere in the direction of its propagation. 

The found directional dependence of accelerations and speeds does not comply with the blast-wave scenario of shock generation in a fairly homogeneous coronal medium. This suggests either that the medium is not homogeneous or that the wave under study is not a blast-generated shock-wave. Although the region to the north of the AR appears to be homogeneous (see \figs{LowCorona}{PFSS}), that is not the case in the vicinity of the AR, as is evident in \fig{xyWaveFronts}. In the case of a piston-driven shock, the shape of the shock wave is determined by the piston shape and the time elapsed until the piston stops acting as shock generator, as well as by the coronal medium.

In a flare-produced shock, a pressure pulse, static and close to the solar surface, would produce a shock of the blast-wave type. The constant speed of the sphere's center used by our model cannot be in agreement with such a shock, even if it is considered that the shock evolves as a freely propagating blast wave after the piston stops acting as the wave driver. 

The hypothesis of a flare impulsive pressure-pulse that generates a blast wave responsible for the Moreton event is unlikely, given that the wave perturbation is seen to start in coincidence with the flare onset time [$t_{on}$] 
(see \fig{Spline}).  It should be noted that the time $t_{on}$ is determined from the same data taken as a basis for the DT maps. 
Following \cite{Vrsnak08}, such coincidence cannot account for the necessary delay in the shock formation after the onset of the flare pressure pulse, given  reasonable values of ambient Alfv\'en speed and piston expansion speed.
Some authors have suggested that the launch of Moreton events precedes the flare impulsive phase \citep{Veronig08, Muhr10}, while others find a close temporal coincidence \citep{Warmuth04b, Temmer09, Francile13}, thus indicating that probably not all Moreton events are similar in this sense. 

To consider the hypothesis of piston-driven shock-wave generation, it is fundamental to understand the characteristics of the potentially associated piston. \citet{Kleint15} measured the acceleration of the filament corresponding to this event, obtaining 3\,--\,5 \kmss~between t\,=\,17:44:30 UT and t\,=\,17:45:00 UT, with the peak upflow velocity at t\,=\,17:45:40 UT. In relation with other event, \cite{Temmer09} found that a synthetic 3D piston of size $\approx$\,110\,Mm accelerated to $\approx$\,4.8\,\kmss~during $\approx$\,160\,s, was able to generate a shock wave in correspondence with the Moreton-wave kinematics. Therefore, the fast ejection of the filament/CME ensemble can be regarded as the piston and thus as responsible for the shock wave generation. It is possible that after the flare's impulsive phase the filament/CME is not impulsively accelerated anymore, and therefore it continues evolving with a nearly constant speed or decelerating, with the shock-wavefront persisting ahead of it. Otherwise, the filament/CME would not be able to generate a shock wave inside the AR, given the high Alfv\'en speed of the region. Only when it reaches regions with low Alfv\'en speed outside the AR, \ie~northwards, it is able to generate a shock wave that steepens and produces the chromospheric and coronal observable signatures \citep{Vrsnak08}. 

Although the model is not accurate enough to draw conclusive results, it yields a good approximation to the scenario of coronal wave generation. If the case of a shock generated by a temporary piston is true, the 3D piston follows an approximate path given by $\vec{v}$=$(v_x, v_y,v_z)$, at least during the investigated time interval. The results of our model have a good match with a sphere that expands at the speed of the fast rising filament measured by \citet{Kleint15} and \citet{Liu15}, suggesting an association between the filament eruption and the wave event.
The piston could thus be attributed to the front of the CME itself, or the full CME bubble or loops, as has been reported by several observational studies \citep{Patsourakos12}. Some observational reports reinforce this picture \citep{Kienreich09, Patsourakos09, Veronig10, Grechnev11b}. If this is the case, the faster lobe of sectors 29\,--\,34 could be related with a piston provided by a CME structure, probably stretched, non-radially rising, \ie~with a northward inclination.

A CME evolving faster than the shock is not a coherent picture, given that the piston should generate and drive the shock. The circular feature indicated by green arrows in \fig{LowCorona} could thus be attributed to a 3D rising structure bent to the north, \eg~an erupting CME bubble driving the shock ahead of it, whereas the observed outermost projected lateral extent of the shock (black arrows) maintains a circular shape.

By assuming that the piston decelerates and stops driving the shock wave at a certain point, as done by several authors \citep[][and references therein]{Patsourakos12}, the shock will become a freely propagating shock wave similar to a blast wave \citep{Warmuth15}, but should prevail with the speed and shape imprinted by the driver at the initial stages. After some time, the shock front would tend to become more homogeneous and would depend on the particular conditions of the ambient corona and interplanetary space. The white-light observations give account of a CME and shock evolution following characteristics similar to those close to the origin.

\subsection{Summary}

We can summarize our main findings for the 29 March 2014 wave-event as follows:
\begin{itemize}
\item{The Moreton event exhibits the typical characteristics as regards speed, deceleration and angular span; 
nevertheless the first H$\alpha$ traces detected are at a relatively short
distance from the brightest flare-kernel site, when compared with
previous statistical results.}
\item{The wave perturbation is seen to appear in coincidence with the estimated flare onset time.}
\item{The analysis by sectors indicates a non-homogeneous angular
behavior of the wavefronts in all analyzed wavelengths.}
\item{Observations in He {\sc ii} 304\,\AA~could be used as tracers and indicators of the existence of H$\alpha$ Moreton-waves even 
if the latter are not observed because of either unfavourable seeing conditions or a lack of ground-based data.}  
\item{The wave speed is higher in the northward direction, which cannot be attributed
to the open magnetic field lines and lower Alfv\'en speed in that region.
This higher speed  could be related to the characteristics of the driver and region where the wave was ignited. However, the first mentioned conditions favor compression over the chromosphere.}
\item{Despite its simplicity, the proposed geometrical model is able to explain the
measured chromospheric traces and the near-surface EUV wave signatures as the intersection of a spherically expanding shock-front with the chromosphere. This spherical shock can be explained as driven by a 3D piston that follows a northward and upward path. The shock trajectory and expansion speed are similar to those of the driver at its initial stages.}
\item{The 3D piston can be attributed to the front of the CME or the
full CME bubble, expanding at the speed of the associated rising
filament.}
\end{itemize}

We plan to apply our simple model to other Moreton events to test the generality of our results. If the model turns out to be feasible, it could be applied to estimate the initial speeds and trajectories of ejected filaments or flux ropes during CME events.

%
\begin{acks}
FML, ML, HC, and CHM acknowledge financial support from grants PICT 2012-0973 (ANPCyT) and PIP 2012-01-403 (CONICET).
CNF acknowledge support from grant PICTO 2009-0177 (FONCYT) and CICITCA-UNSJ E936.
ML and CHM recognize grant UBACyT 20020130100321.
HC and FML appreciate support from project UTI1744 (UTN). HC and CHM are members of the Carrera del Investigador Cient\'ifico (CONICET). FML is a fellow of CONICET. ML is a member of the Carrera del Personal de Apoyo (CONICET). DML is an Early Career Fellow funded by the Leverhulme Trust. We thank the reviewer of our manuscript for his/her constructive comments and suggestions, 
\end{acks}

\section*{Disclosure of Potential Conflicts of Interest}
The authors declare that they have no conflicts of interest.

 \bibliographystyle{spr-mp-sola-cnd}
 \bibliography{moreton-bib}

\end{article}
\end{document}